\documentclass[pdflatex, sn-mathphys-ay, iicol]{sn-jnl}

\usepackage{graphicx} %
\usepackage{subcaption} %
\usepackage{multirow} %
\usepackage{amsmath, amssymb, amsfonts, amsthm} %
\usepackage{mathrsfs}%
\usepackage[title]{appendix}%
\usepackage[dvipsnames]{xcolor}
\usepackage{textcomp}%
\usepackage{manyfoot}%
\usepackage{booktabs}%
\usepackage{algorithm, algorithmicx, algpseudocode, listings} %
\usepackage[numbered,framed]{matlab-prettifier}
\usepackage{tikz}
    \usetikzlibrary{patterns}
    \usetikzlibrary{math}
    \usetikzlibrary{calc}
    \usetikzlibrary{arrows.meta}
\usepackage{pgfplots}
    \pgfplotsset{compat=1.18}
\usepackage{verbatim}

\definecolor{LightCyan}{rgb}{0.88,1,1}
\definecolor{LightCyan}{rgb}{0.88,1,1}
\definecolor{mygreen}{RGB}{28,172,0} 
\definecolor{mylilas}{RGB}{170,55,241}
\usepackage{etoolbox}

\begin{document}

\title[TMC matlab code]{A Matlab code for analysis and topology optimization with Third Medium Contact}

\author[1]{\fnm{Andreas Henrik} \sur{Frederiksen}}\email{andfr@dtu.dk}

\author[1]{\fnm{Ole}
\sur{Sigmund}}\email{olsi@dtu.dk}

\author*[1]{\fnm{Federico}
\sur{Ferrari}}\email{feferr@dtu.dk}

\affil[1]{\orgdiv{Section of Solid Mechanics, Department of Civil and Mechanical Engineering},
\orgname{Technical University of Denmark (DTU)},
\orgaddress{\street{Koppels Allé, 404},
\city{Kgs. Lyngby}, \postcode{2800}, \country{Denmark}}}

\abstract{We present a \textsf{Matlab} code for modelling and topology optimization of hyperelastic structures, including contact modelled by the Third Medium Contact (TMC) approach. By using the so-called \textsf{HuHu}-regularization we penalize the skew distortion of the bilinear finite elements discretizing void regions, thus promoting convergence of the nonlinear solver. First, we show how this method is implemented in a compact code, allowing to simulate contact and force transfer in hyperelastic structures. We then solve two topology optimization problems for minimum end-compliance of structures exhibiting contact. In the first example, contact happens at the supported boundary, while the second features self-contact. The \textsf{Matlab} scripts that replicate the results are included, and we discuss some possible extensions to more general problems.}

\keywords{Third Medium Contact, Topology Optimization, Hyperelasticity, Matlab}

\maketitle

\section{Introduction}
 \label{Sec:Introduction}

This work provides a user-friendly, open-source \textsf{Matlab} implementation of the Third Medium Contact (TMC) model, and of its application to density-based Topology Optimization (TO).

Contact simulation by the TMC method was initiated by \cite{wriggers-etal_13a_thirdMediumContact} and \cite{bog-etal_15a_normalContactFIctitiousMaterial}; however, convergence issues due to the extreme deformation of finite elements (FE) discretizing the void regions prevented the method from taking hold. Recently, TMC has been revived in the context of topology optimization (TO) by \cite{bluhm-etal_21a_internalContactModelingTO}, thanks to the introduction of a numerical stabilization method, later coined ``\textsf{HuHu}-regularization'', which cured the aforementioned convergence issues.

Since then, the synergistic use of TMC within TO has found increasing applications, including the design of self-contacting structures undergoing large deformations \citep{frederiksen-etal_24a_topOptSelfContacting}, metamaterials \citep{dalklint-etal_23a_computationalSelfContact}, thermal switches \citep{dalklint-etal_25a_topOptThermoRegulators}, 3D contacting hooks \citep{frederiksen-etal_25a_inprovedContact3DTO}, pneumatic actuators for robotics \citep{mehta-etal_25a_toopOptSoftActuators}, springs \citep{bluhm-etal_23a_inverseContactMechanicaSprings}, and multi-stable lattices showing snap-through \citep{aveline-etal_25a_inverseDesignMultiscaleLattices}.

The plethora of applications clearly shows the relevance and versatility of TMC for the design of advanced structures and materials taking advantage of contact response. Contrary to classical contact formulations \citep{book:wriggers_2006, deLorenzis-etal_17a_computationalContactMechanics}, the TMC approach is fully implicit, avoiding the need for explicit tracking of the contact interfaces, and the associated Lagrange multipliers. This allows its seamless integration in optimization frameworks, since the small but finite stiffness of void regions, represented by the so-called \emph{Third Medium}, provide a differentiable contact response. In particular, TMC is well-suited for use within TO, since it allows the contact interface to form naturally within the design domain, thus preserving full design freedom. The only other fully implicit family of contact formulation is based on phase-field methods \citep{lorez-etal_24a_eulerianContactPhaseField, lorez-pundir_25a_frictionContactEulerian}, but has to date not been tested within TO.

The spreading of TO in academia and industry greatly benefits from educational codes, which facilitate learning of basic TO concepts \citep{sigmund_01a_99LinesTopOptCode}, but also offer an assisted introduction to more advanced methods and applications \citep{ferrari-etal_21a_250LinesTopOptBuckling, giraldolondono-paulino_21a_polyStress, woldseth-etal_24a_808phasorDehomogenization}. The number of educational codes, largely written in \textsf{Matlab} and addressed at various physics or applications, is steadily growing and we refer to \cite{wang-etal_21a_comprehensiveReviewEducationalCodes} for a thorough list of contributions.

The recent improvement and revival of TMC, and its application to TO offers a timely opportunity for presenting yet another educational code. The present \textsf{Matlab} implementation builds on the framework and methods established in \cite{andreassen-etal_11a_88LinesTopOpt} and \cite{ferrari-sigmund_20b_99LinesNewGeneration}. Thus, the user is supposed to be familiar with those codes and with the basics of density-based TO \citep{book:bendsoe-sigmund_2004}.

The code has been written for compactness, readability, and modularity, rather than aiming at optimal efficiency or general robustness. The implementation has been split into five subroutines, such that the advanced user may adapt or extend each of them to handle more general problems, elements and regularizations \citep{wriggers-etal_25a_tmcFirstSecondElement, wriggers-etal_25a_lowOrderElementsTMC}, or heavier computational tasks. Two main scripts are included, which can be used for replicating the results shown in the paper.

We believe that this can be a relevant contribution to the field, showing the inherent simplicity of implementation and effectiveness of TMC for solving complex topology design problems of  hyperelastic structures in contact. To date, this is the only code available covering such advanced task. A formulation for TMC for contact analysis (without optimization) is currently available in the \textsf{AceFEM} software \citep{korelc_aceFEM, wriggers-etal_25a_tmcFirstSecondElement}. The recently published work by \cite{bin-etal_25a_499FrictionalContactTO} introduces a \textsf{Matlab} code for solving TO problems with contact, limited to linear elastic response and using a classical, explicit contact formulation.

This paper is organized as follows. \autoref{Sec:modelingTMC} gives an overview of contact modelling by the TMC method and the \textsf{HuHu}-regularization. A popular benchmark example, which can be replicated by the script \texttt{cshapeTMC.m}, is shown in \autoref{sSec:analysisCshape}. \autoref{Sec:TOframework} introduces the density-based TO framework used in the code, and \autoref{sSec:topologyOptimizationExample} shows an example involving minimum end-compliance design, taking advantage of contact, which can be replicated by the script \texttt{topTMC.m}. Finally, in \autoref{Sec:disussionConclusions} we discuss possible extensions to the code provided.

The complete \textsf{Matlab} code is listed in \autoref{Sec:matlabImplementation}, where we discuss in detail the implementation, its underlying assumptions and input data, and we give some specific hints on extensions for generality or improved efficiency.

\subsection{Conventions and notation}
 \label{sSec:notationConventions}

We use italic for the continuum formulation, with rank-1 and rank-2 tensors denoted by lower- and upper-case boldface letters (e.g., $\boldsymbol{a}$, $\boldsymbol{A}$). In the discretized setting we use Roman letters, with the same convention for vectors and matrices (e.g., $\mathbf{a}$, $\mathbf{A}$). The determinant of a tensor (viz. matrix) is denoted as $|\boldsymbol{A}|$, (viz. $|\mathbf{A}|$). The contraction of two $n$-th order tensors is denoted by $\boldsymbol{A}\cdot\boldsymbol{B} = A_{i,j,\ldots, n}B_{i,j,\ldots, n}$, with contraction symbol ``$\cdot$'', and implying summation over repeated indices (Einstein's convention).

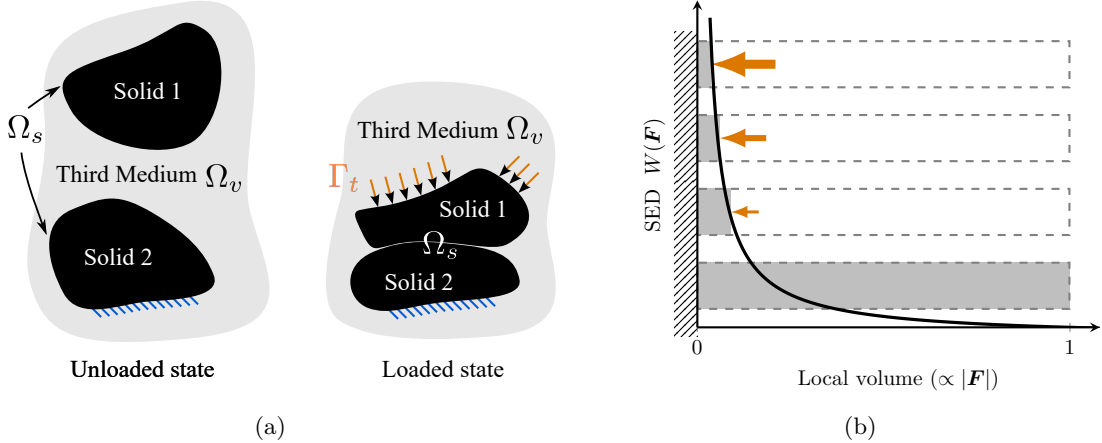
\begin{figure*}
    \centering
    \subfloat[]{
    \includegraphics[scale = 0.9,keepaspectratio]
    {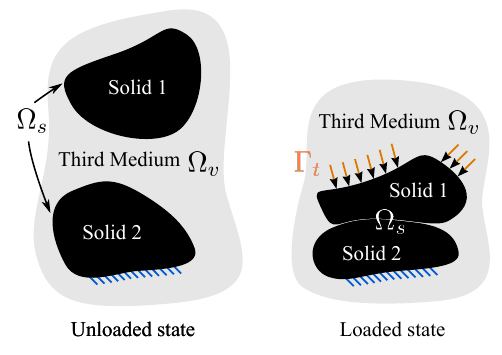}} \qquad
    \subfloat[]{
    \begin{tikzpicture}[scale = 0.8]
    \tikzmath{\x1 = 0.035; \x2 = 0.06; \x3 = 0.09;}
    \definecolor{gscolor}{RGB}{220,120, 0} 
    \begin{scope}[shift={(0,0)}, scale=1.53]
    \fill [pattern=north east lines] (-0.25,-0.1) rectangle (0,3.2);
    \draw [black, thick, dashed, fill=none, opacity=0.5] (0,.2) rectangle (4,.7);
    \draw [black, thick, dashed, fill=none, opacity=0.5] (0,1.0) rectangle (4,1.5);
    \draw [black, thick, dashed, fill=none, opacity=0.5] (0,1.8) rectangle (4,2.3);
    \draw [black, thick, dashed, fill=none, opacity=0.5] (0,2.6) rectangle (4,3.1);
    \fill [gray, opacity=0.5] (0,.2) rectangle (4,.7);
    \fill [gray, opacity=0.5] (0,1.0) rectangle ({\x3*4},1.5);
    \fill [gray, opacity=0.5] (0,1.8) rectangle ({\x2*4},2.3);
    \fill [gray, opacity=0.5] (0,2.6) rectangle ({\x1*4},3.1);
    \draw [-latex, gscolor, line width=0.4mm] ({\x3*4+0.3},1.25) -- ({\x3*4},1.25);
    \draw [-latex, gscolor, line width=0.8mm] ({\x2*4+0.5},2.05) -- ({\x2*4},2.05);
    \draw [-latex, gscolor, line width=1.2mm] ({\x1*4+0.7},2.85) -- ({\x1*4},2.85);
    \end{scope}
    \begin{axis}[
        axis lines = left,
        xlabel =  {Local volume ($\propto |\boldsymbol{F}|$)},
        ylabel = {SED $\ W(\boldsymbol{F})$},
        width = 8.2cm,
        height = 7cm,
        ytick=\empty,
        ylabel style={yshift=9pt, xshift=-5pt},
        xmin=0,xmax=1.08,
        ymin=0,ymax=30,
        xtick ={0, 1},
        line width=1pt,
    ]
    \addplot [
        domain=0.034:1.08, 
        samples=200, 
        color=black,
        line width=1.5pt,
    ]
    {1/x-1};
    \end{axis}
    \end{tikzpicture}
    }
    \vspace{0mm}
    \caption{\small{(a) Illustration of two solid bodies, collectively occupying the region $\Omega_{s}$, interacting through the Third Medium (TM), which fills the region $\Omega_{v}$. (b) Qualitative illustration of the stiffening behaviour of the material law \eqref{eq:sedNeoHookean}, as the local volume shrinks to zero under uniaxial compression. The arrows show the normal force for a reference volume (dashed) compressed to a current volume (gray) for higher compression levels}}
    \label{fig:TMCconcept}
\end{figure*}

\section{Contact modelling by TMC}
 \label{Sec:modelingTMC}

The key idea in TMC modelling is illustrated in \autoref{fig:TMCconcept}. The void region surrounding the two solids coming into contact is modelled as a nonlinear material by introducing a \textit{fictitious} Third Medium (TM) featuring: (1) very low, \emph{but finite} stiffness at small deformations; (2) exponentially increasing stiffness as the local volume shrinks to zero (see \autoref{fig:TMCconcept}(b)).

These two features give the following response, as the solids come into contact:
\begin{itemize}
 \item a small \emph{but finite} force is transferred between the two solids even at long range, before contact is engaged;
 \item the TM acts as a barrier, preventing the solid bodies from penetrating each other as the contact is established.
\end{itemize}

These features make the TMC very attractive for contact modelling within TO, avoiding the need for methods tracking the solid/void interface and explicit modelling of force transfer by nonlinear constraints, ultimately making the optimization problem non-differentiable \citep{book:wriggers_2006, deLorenzis-etal_17a_computationalContactMechanics}. Most importantly, without the TMC the topology optimizer would have no means to discover the advantage of establishing contact within the design domain.

The TM is assigned a neo-Hookean material law \citep{book:holzapfel} with Strain Energy Density (SED)
\begin{equation}
 \label{eq:sedNeoHookean}
 \begin{aligned}
  W(\boldsymbol{u}) & = \dfrac{\lambda}{2} \ln(|\boldsymbol{F}|)^2 + \\
  & \dfrac{\mu}{2} \left(\mathrm{tr}(\boldsymbol{F}^{T}\boldsymbol{F}) - 3\right) - \mu \ln(|\boldsymbol{F}|)
 \end{aligned}
\end{equation}
where $\lambda$, $\mu$ are the initial Lamé parameters, and $\boldsymbol{F} = \boldsymbol{F}(\boldsymbol{u}) = \boldsymbol{I} + \nabla \boldsymbol{u}$ is the deformation gradient.

The first term in \eqref{eq:sedNeoHookean}, proportional to $\ln(|\boldsymbol{F}|)^2$, accounts for volumetric changes, and is pivotal for providing the TM with the extreme stiffening response depicted in \autoref{fig:TMCconcept} (b). Clearly, the TM can also be modelled by other hyperelastic material laws accommodating large deformations and showing the same exponential stiffening for $|\boldsymbol{F}| \rightarrow 0$, such as those listed in \citet{klarbring-stromberg_13a_hyperelasticTO, dalklint-etal_23a_computationalSelfContact}. 

We point out that by using TMC, a small gap between the solid interfaces still exists, even when full contact is considered (see \autoref{fig:TMCconcept}(a)). For practical applications, the material parameters of the third medium are chosen such that this small gap cannot be visually observed, and has no practical influence on the systems' response.

For simplicity, and following \cite{frederiksen-etal_25a_inprovedContact3DTO}, we use the same neo-Hookean material for both the solid and the TM, and the SED \eqref{eq:sedNeoHookean} is scaled by the factor $k_{v}\ll 1$ in the TM region, making it multiple orders of magnitude softer compared to the solid.

The nonlinear equilibrium equation follows from the first variation of the total potential energy (TPE)
\begin{equation}
 \label{eq:variationTPE}
  \begin{aligned}
  & \delta\Pi(\Lambda, \boldsymbol{u};\delta \boldsymbol{u}) = 
  \mathcal{R}_{\rm int}(\boldsymbol{u};\delta \boldsymbol{u}) - \Lambda\ell(\delta\boldsymbol{u}) = \\ &
  \int_{\Omega_{s}\cup\Omega_{v}} \delta W(\boldsymbol{u};\delta\boldsymbol{u}) \: \mathrm{d}\Omega
  - \Lambda\ell(\delta\boldsymbol{u}) = 0 \quad \forall\: \delta\boldsymbol{u}
 \end{aligned}
\end{equation}
where $\mathcal{R}_{\rm int}$ are the internal forces, and the loading is assumed proportional to $\Lambda\in\mathbb{R}$. The linear form associated with the surface tractions $\bar{t}$, reads
\begin{equation}
 \label{eq:variationLF}
  \ell(\delta\boldsymbol{u}) = \int_{\Gamma_{t}}
  \bar{t} \cdot \delta\boldsymbol{u} \: \mathrm{d}\Gamma_{t}
\end{equation}
whereas the variation of the SED \eqref{eq:sedNeoHookean} gives
\begin{equation}
 \label{eq:variationSED}
  \delta W(\boldsymbol{u};\delta\boldsymbol{u}) = \boldsymbol{P}\cdot\delta\boldsymbol{F}
\end{equation}
where $\boldsymbol{P} = \lambda\ln(|\boldsymbol{F}|)\boldsymbol{F}^{-T} +\mu(\boldsymbol{F}-\boldsymbol{F}^{-T})$ is the first Piola-Kirchhoff stress tensor.

\begin{figure}
 \centering
  \includegraphics[width=\linewidth]{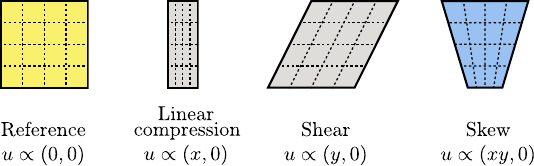}
  \caption{\small{Illustrations of the deformations that can be represented by bilinear $\mathcal{Q}_{1}$ elements. The \textsf{HuHu} regularization only penalizes the skew deformation}}
 \label{fig:huhuRegularization}
\end{figure}

\subsection{\textsf{HuHu}-regularization}
 \label{sSec:huhuRegularization}

As contact is approached some compressed, heavily distorted regions in the TM jeopardize the stability of Equation \eqref{eq:variationTPE}, which is discretized by finite elements (FE). Therefore, some form of stabilization is necessary to achieve convergence of the nonlinear equilibrium solver. 

The revival of TMC within TO was indeed driven by a regularization method proposed by \cite{bluhm-etal_21a_internalContactModelingTO}, and then popularized and extended by \cite{frederiksen-etal_24a_topOptSelfContacting, TMCwebpageWikipedia} under the name of \textsf{HuHu}-regularization.

The idea of \textsf{HuHu}-regularization is to penalize specific high order, skew deformations of the FE discretizing the TM, such that these do not become extreme, cause element inversion, and thereby destroy convergence. The bilinear, $\mathcal{Q}_{1}$ element used in this work allows deformations spanned by the modes shown in \autoref{fig:huhuRegularization}. In this case, the \textsf{HuHu}-regularization only penalizes the skew deformation, thus leaving linear compression and shear deformations unaffected.

The \textsf{HuHu} regularization is introduced by augmenting the SED \eqref{eq:sedNeoHookean} as
\begin{equation}
 \label{eq:huhuRegularization}
  \tilde{W}(\boldsymbol{u}) = W(\boldsymbol{u}) +
  \dfrac{k_{r}}{2}\mathbb{H}\boldsymbol{u}
  \cdot\mathbb{H}\boldsymbol{u}
\end{equation}
where $\mathbb{H}\boldsymbol{u} := \frac{\partial^{2} u_{i}}{\partial X_{i}\partial X_{j}}$ is the spatial Hessian of the displacement field.

The scaling factor $k_{r} = \alpha L^2 (K+\frac{4}{3}G)$, depends on a characteristic domain length, $L$, the coefficient $\alpha > 0$, which is set in the range $\alpha \in [10^{-5}, 10^{-7}]$ \citep{frederiksen-etal_24a_topOptSelfContacting, frederiksen-etal_25a_inprovedContact3DTO} and the initial bulk and shear moduli $(K,G)$ of the solid material.

\begin{figure}[t]
  \centering
   \begin{tikzpicture}[scale=1.0]
     \begin{axis}[
      width=8cm,
      height=5cm,
      axis lines=middle,
      xlabel={$\nu$},
      ylabel={},
      xmin=-1, xmax=0.6,
      ymin=0, ymax=4,
      xtick={-1,-0.5,0.0,0.2,0.5},
      ytick={1,2,3},
      legend style={at={(0.,1.0)},anchor=north west},
      domain=-0.98:0.48,
      samples=200,
      thick,
      axis line style={-Stealth},
      y axis line style={-Stealth},
      x axis line style={-Stealth},
      axis x line = middle,
      axis y line = middle]
      \definecolor{customorange}{RGB}{220,120,0}
      \addplot[color=customorange, solid, very thick] {1/(3*(1-2*x))};
      \addlegendentry{$K$};
      \addplot[RoyalPurple, dashed, very thick] {( 1/(3*(1-2*x)) + (4/3)*1/(2*(1+x)))};
      \addlegendentry{$K + \frac{4}{3}G$};
      \end{axis}
    \end{tikzpicture}
 \caption{\small{Bulk modulus $K$, and longitudinal modulus $M = K + \frac{4}{3}G$ as functions of the Poisson's ratio $\nu$ ($E=1$). The two moduli become coincident for $\nu \rightarrow 0.5$. However, for low $\nu$ values $K < 1$, whereas $M$ is always $> 1$. We remark that all material coefficients must be considered as \textit{initial} elastic moduli, such that \eqref{eq:sedNeoHookean} will reduce to the linear model as $\boldsymbol{F}\rightarrow\boldsymbol{I}$}}
 \label{fig:modulusComparison}
\end{figure}
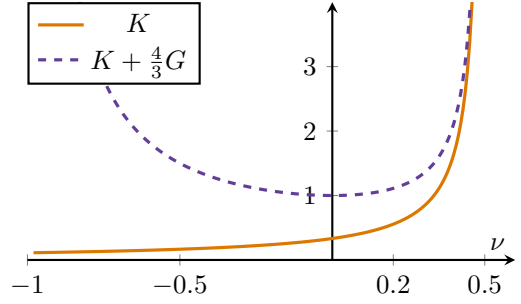

We note that in previous works $k_{r}$ was a function of the bulk modulus only \citep{frederiksen-etal_24a_topOptSelfContacting,frederiksen-etal_25a_inprovedContact3DTO, bluhm-etal_21a_internalContactModelingTO,dalklint-etal_25a_topOptThermoRegulators}, which made the regularization essentially insensitive to the value of the shear modulus ($G$). A more robust, and physically sound alternative is to make $\alpha$ proportional to the longitudinal modulus $M = K + \frac{4}{3}G$ \citep{book:holzapfel}, also known as the $P$-wave modulus, since the TM regions near contact interfaces experience uniaxial compression, rather than pure volumetric change. Thus, using $M$ appears more rational than just $K$, with the added educational benefit of keeping the regularization material parameter $>1$ also for low values of the Poisson's ratio $\nu$ (see \autoref{fig:modulusComparison}).

The variation of the augmented SED, introducing the regularization \eqref{eq:huhuRegularization} reads
\begin{equation}
 \label{eq:variationSED_HuHu}
  \delta\tilde{W}(\boldsymbol{u};\delta\boldsymbol{u}) =
   \boldsymbol{P}\cdot\delta\boldsymbol{F} 
   + k_{r} e^{-5|\boldsymbol{F}|}
   \mathbb{H}\boldsymbol{u}
   \cdot \mathbb{H}\delta\boldsymbol{u}
\end{equation}
where the scaling factor $e^{-5|\boldsymbol{F}|}$, which is not explicitly included in the SED \citep{bluhm-etal_21a_internalContactModelingTO}, reduces the regularization effect on elements that are not heavily compressed, and generally makes the whole tangent matrix non-symmetric.

Details on the discretization of \eqref{eq:variationLF} and \eqref{eq:variationSED_HuHu}, and on the solution of \eqref{eq:variationTPE} by an incremental iterative method are given in \ref{sApp:incrementalIterativeSolver} and \ref{sApp:assemblyKtFi}.

We remark that some alternative formulations for the regularization of the TMC approach exist \citep{faltus-etal_24a_tmcPneumaticallyActuated, wriggers-etal_25a_tmcFirstSecondElement}, and some may also be more effective than the \textsf{HuHu}, when tested on low-order finite elements. We refer to \citet{weissenfels-etal_15a_contactLayerLargeDeformations, wriggers-etal_25a_tmcFirstSecondElement} for a comparison of some different TMC regularization formulations. Here we adopt the \textsf{HuHu} due to its ease of implementation and acceptable robustness.

For completeness, we also recall the \textsf{HuHu-LuLu}-regularization, proposed by \cite{frederiksen-etal_25a_inprovedContact3DTO}. Making use of \eqref{eq:huhuRegularization}, this improved regularization reads
\begin{equation}
 \label{eq:huhululuRegularization}
 \tilde{\tilde{W}}(\boldsymbol{u}) = \tilde{W}(\boldsymbol{u}) - \frac{k_{r}}{2{\rm tr}(\boldsymbol{I})} \mathbb{L}\boldsymbol{u}\cdot \mathbb{L}\boldsymbol{u}
\end{equation}
where $\mathbb{L}\boldsymbol{u} := \frac{\partial^{2}u_{i}}{\partial X_{j}\partial X_{j}}$ is the Laplacian of the displacement field. 

The added term, proportional to $\mathbb{L}\boldsymbol{u}\cdot \mathbb{L}\boldsymbol{u}$, reduces the penalization of nonlinear compression and bending modes, which carry important physical information in higher-order elements. However, for the bilinear $\mathcal{Q}_{1}$ element used here we have $\mathbb{L}\boldsymbol{u} = 0$, and therefore \eqref{eq:huhuRegularization} and \eqref{eq:huhululuRegularization} are identical.

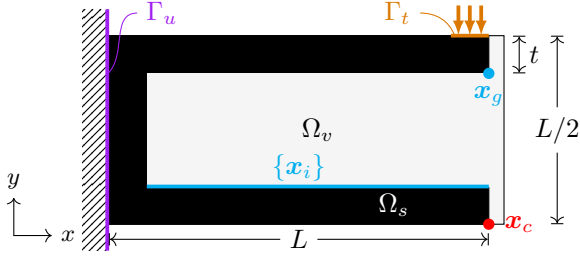
\begin{figure}[tb]
 \centering
  \begin{tikzpicture}
   \definecolor{customgray}{RGB}{245, 245, 245}
   \definecolor{custompurple}{RGB}{160, 32, 240}
   \definecolor{customorange}{RGB}{220, 120, 0}
   \draw [fill=customgray, draw=black] (0.,0.) rectangle (5.2,2.5);
   \draw [fill=black, draw=black] (0.,0.) rectangle (5.0,2.5);
   \draw [fill=customgray, draw=none] (0.5,0.5) rectangle (5.1,2);
   \fill [pattern=north east lines] (-0.35,-0.35) rectangle (-0.02,2.85);
   \draw [color=custompurple, line width=0.5mm] (-0.02,-0.35) -- (-0.02,2.85);
   \node (L) at (2.5,-0.2) {\normalsize $L$};
   \draw [|<-] (0,-0.2) -- (L);
   \draw [->|] (L) -- (5,-0.2);
   \node (H) at (5.9, 1.25) {\normalsize $L/2$};
   \draw [|<-] (5.9, 0) -- (H);
   \draw [->|] (H) -- (5.9,2.5);
   \node (t) at (5.6, 2.25) {\normalsize $t$};
   \draw [|<->|] (5.4, 2.0) -- (5.4,2.5);
   \node [color=white] (omega_s) at (3.75, 0.25) {\normalsize $\Omega_s$};
   \node (omega_v) at (2.75, 1.25) {\normalsize $\Omega_v$};
   \node [color=customorange] (gamma_t) at (3.8, 2.8) {\normalsize $\Gamma_{\scriptstyle t}$};
   \node [color=custompurple] (gamma_u) at (0.7, 2.8) {\normalsize $\Gamma_{\scriptstyle u}$};
   \draw [color=custompurple] (gamma_u.west) to[out=180, in=45] (-0.01,2.0);
   \draw [color=customorange] (gamma_t.east) to[out=0, in=110] (4.55,2.5);
   \draw [color=customorange, line width=0.5mm] (4.5,2.5) -- (5,2.5);
   \node at (5,2) [circle, fill, color=cyan, inner sep=1.5pt]{};
   \node [color=cyan] (x_g) at (5, 1.7) {\normalsize $\boldsymbol{x}_{g}$};
   \node at (5,0) [circle, fill, color=red, inner sep=1.5pt]{};
   \node [color=red] (x_c) at (5.4, 0.0) {\normalsize $\boldsymbol{x}_{c}$};
   \draw [color=cyan, line width=0.5mm] (0.5,0.5) -- (5,0.5);
   \node [color=cyan] (x_i) at (2.5, 0.75) {\normalsize $\{\boldsymbol{x}_{i}\}$};
   \foreach \x in {4.6, 4.75, 4.9}{
    \draw [color=customorange, line width=0.5mm, -{Latex[length=2mm, width=1.5mm]}]
    (\x,2.9) -- (\x,2.5);
    }
   \draw [<->] (-1.25,0.35) [anchor=south] node {\scriptsize $y$} -- (-1.25,-0.15) -- (-0.75,-0.15)  [anchor=west] node {\scriptsize $x$};
  \end{tikzpicture}
  \caption{\small{Geometry and mechanical setup for the C-shape example. The thickness of the solid region $\Omega_{s}$ is $t = 0.1L$, and the void region $\Omega_{v}$ extends of $t/2$ to the right of the solid part}}
 \label{fig:exampleCshapeSetup}
\end{figure}

\subsection{C-shape example}
 \label{sSec:analysisCshape}

We consider the configuration in \autoref{fig:exampleCshapeSetup}, which is a well-established benchmark for testing the regularization of void elements undergoing large deformations \citep{yoon-etal_05a_elementConnectivityParametrization, wang-etal_14a_interpolationLargeStrain}, and recently also used for contact analysis by TMC \citep{bluhm-etal_21a_internalContactModelingTO, faltus-etal_24a_tmcPneumaticallyActuated, wriggers-etal_25a_tmcFirstSecondElement}.

The neo-Hookean material law \eqref{eq:sedNeoHookean}, with Lam\'{e} parameters $\lambda \approx 57.692$ MPa, $\mu \approx 38.462$ MPa, corresponding to the user-defined values $E_{0} = 100$ MPa and $\nu = 0.3$, is assigned to the whole domain $\Omega_{s} \cup \Omega_{v}$, and scaled by $k_{v} = 10^{-6}$ in the TM region $\Omega_{v}$. The \textsf{HuHu} stabilization term is weighted by $\alpha = 10^{-6}$. The left edge is fixed ($\Gamma_{\boldsymbol{u}}$), and a downward uniform traction, with end magnitude $|q| = 30 \: {\rm N}$ is applied over the region $\Gamma_{t}$, extending by $t$ to the left of the top-right solid corner. The user-defined parameters used for obtaining the following results are listed in \autoref{tab:parametersCshape}.

\begin{table}[t]
    \centering
    \caption{\small{Input parameters to the script \texttt{cshapeTMC.m}.}}
    \label{tab:parametersCshape}
    \renewcommand*{\arraystretch}{1.0}
    \begin{tabular}{llll}
     \toprule
      Parameter & Code symbol(s) & Value(s) & Unit \\
      \midrule
      Domain lengths            & \texttt{[Lx,Ly]}       & $[100,50]$           & mm \\
      Solid thickness           & \texttt{thk}           & $10$                 & mm \\
      Young's modulus           & \texttt{E0}            & $100$                & MPa \\
      TMC contrast              & \texttt{kv}            & $10^{-6}$            & -- \\
      Poisson's ratio           & \texttt{nu}            & $0.3$                & -- \\
      \textsf{HuHu} parameter   & \texttt{alpha}         & $10^{-6}$            & -- \\
      \# of elements            & \texttt{[nelx,nely]}   & $[62,30]$            & -- \\
      End load multiplier       & \texttt{lambdaMax}     & $3\cdot10^{-2}E_{0}$ & MPa \\
      \# of load steps          & \texttt{nIncr}         & $200$                & -- \\
      convergence tol.          & \texttt{tolRelRes}     & $10^{-6}$            & -- \\
      \# of Newton iters.       & \texttt{maxIter}       & $50$                 & -- \\
     \bottomrule
    \end{tabular}
\end{table}

The analysis is performed by a simple load-controlled incremental-iterative process, using a Total Lagrangian description and the Green-Lagrange strains \citep{book:crisfield91}. For each load step, governed by the parameter $\Lambda\in [0,1]$ the equilibrium displacement is computed by a Newton iteration (see \autoref{sApp:incrementalIterativeSolver} for details).

\begin{figure*}
  \centering
   \begin{subfigure}[b]{0.3\textwidth}
    \begin{tikzpicture}
     \node[anchor = north west, inner sep=0](img) at (0,2.75) {\includegraphics[clip,scale = 0.265]
     {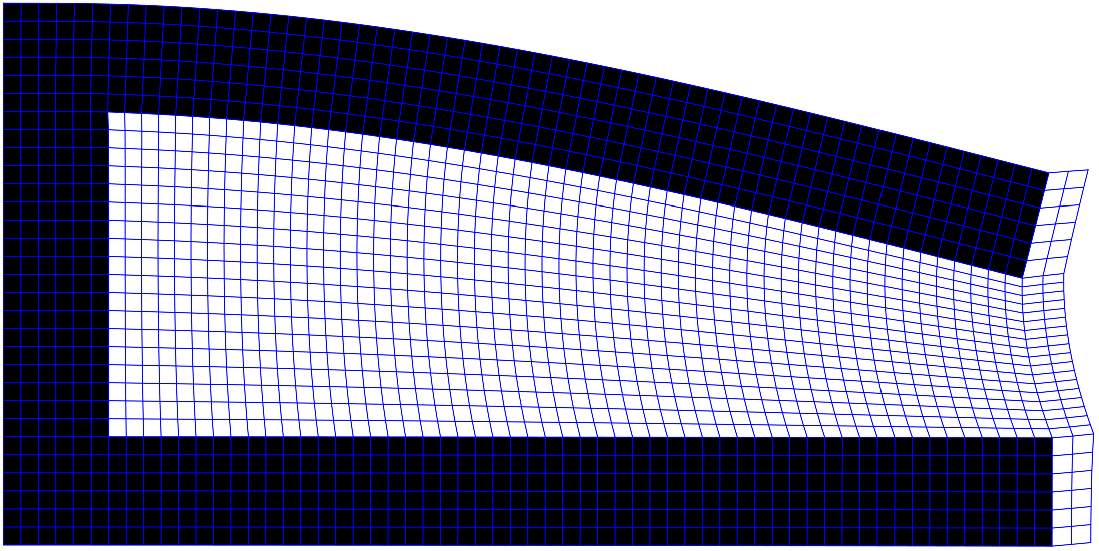}};
     \fill [pattern=north east lines] (-0.35,-0.35) rectangle (0,3.2);
     \draw [-] (0,-0.35) -- (0,3.2);
    \end{tikzpicture}
    \begin{tikzpicture}
     \node[anchor = north west, inner sep=0](img) at (0,2.75) {\includegraphics[clip,scale = 0.265]
     {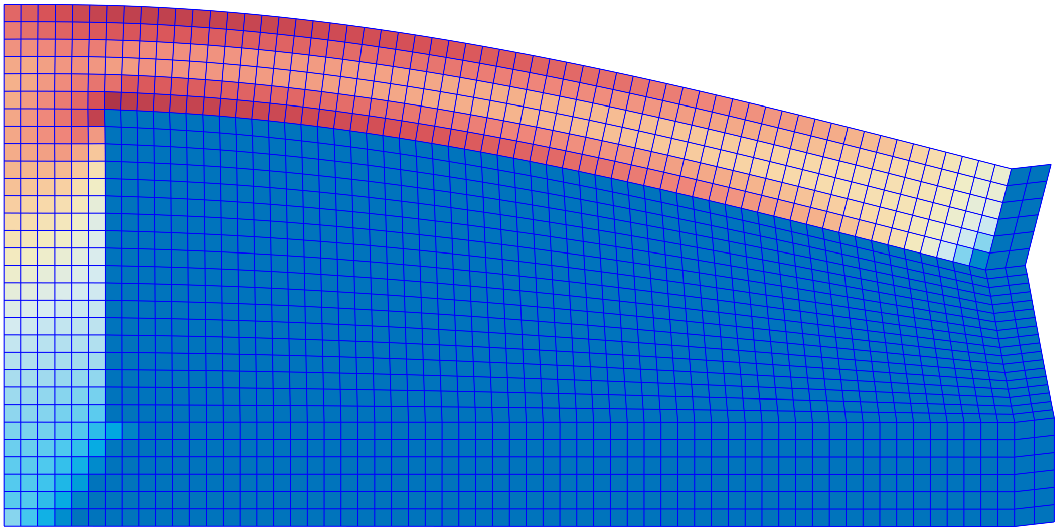}};
     \fill [pattern=north east lines] (-0.35,-0.35) rectangle (0,3.2);
     \draw [-] (0,-0.35) -- (0,3.2);
    \end{tikzpicture}
    \caption{$\Lambda = 0.2$}
    \label{fig:C_shape_a}
   \end{subfigure}
 \hfill
   \begin{subfigure}[b]{0.3\textwidth}
    \begin{tikzpicture}
     \node[anchor = north west, inner sep=0](img) at (0,2.75) {\includegraphics[clip,scale = 0.265]
     {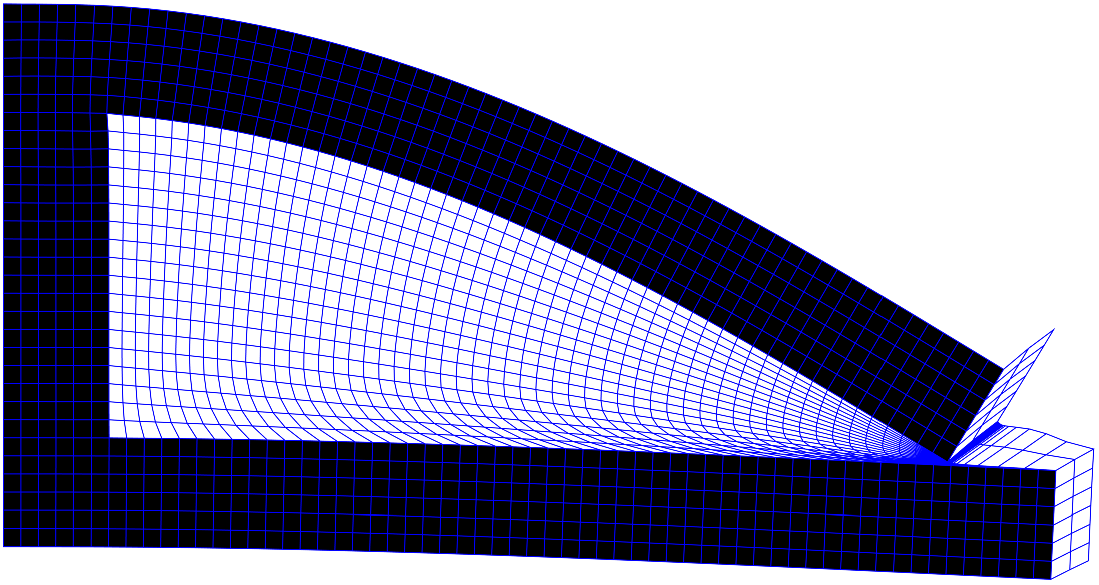}};
     \fill [pattern=north east lines] (-0.35,-0.35) rectangle (0,3.2);
     \draw [-] (0,-0.35) -- (0,3.2);
    \end{tikzpicture}
    \begin{tikzpicture}
     \node[anchor = north west, inner sep=0](img) at (0,2.75) {\includegraphics[clip,scale = 0.265]
     {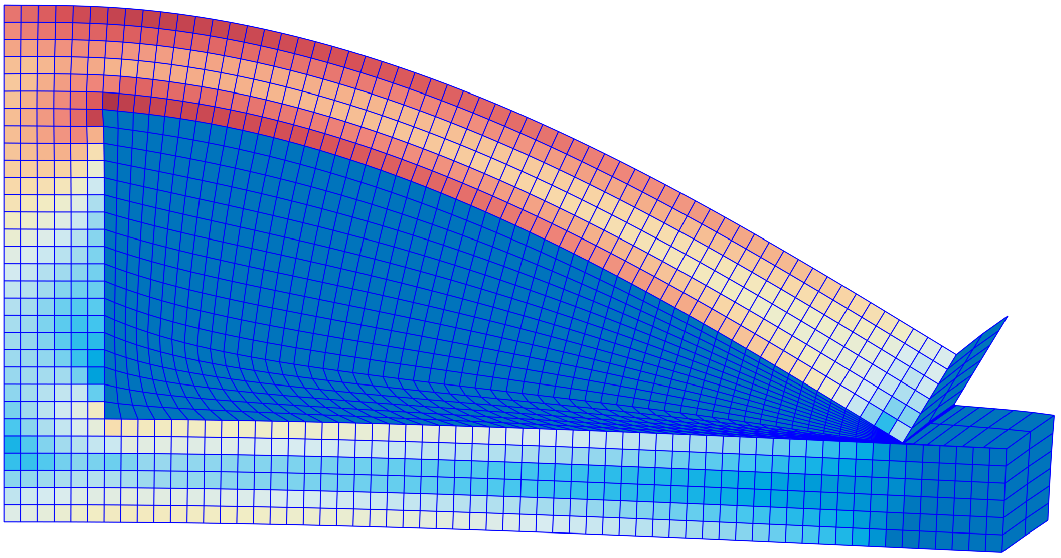}};
     \fill [pattern=north east lines] (-0.35,-0.35) rectangle (0,3.2);
     \draw [-] (0,-0.35) -- (0,3.2);
    \end{tikzpicture}
    \caption{$\Lambda = 0.5$}
    \label{fig:C_shape_b}
   \end{subfigure}
 \hfill
   \begin{subfigure}[b]{0.3\textwidth}
    \begin{tikzpicture}
     \node[anchor = north west, inner sep=0](img) at (0,2.75) {\includegraphics[clip,scale = 0.265]
     {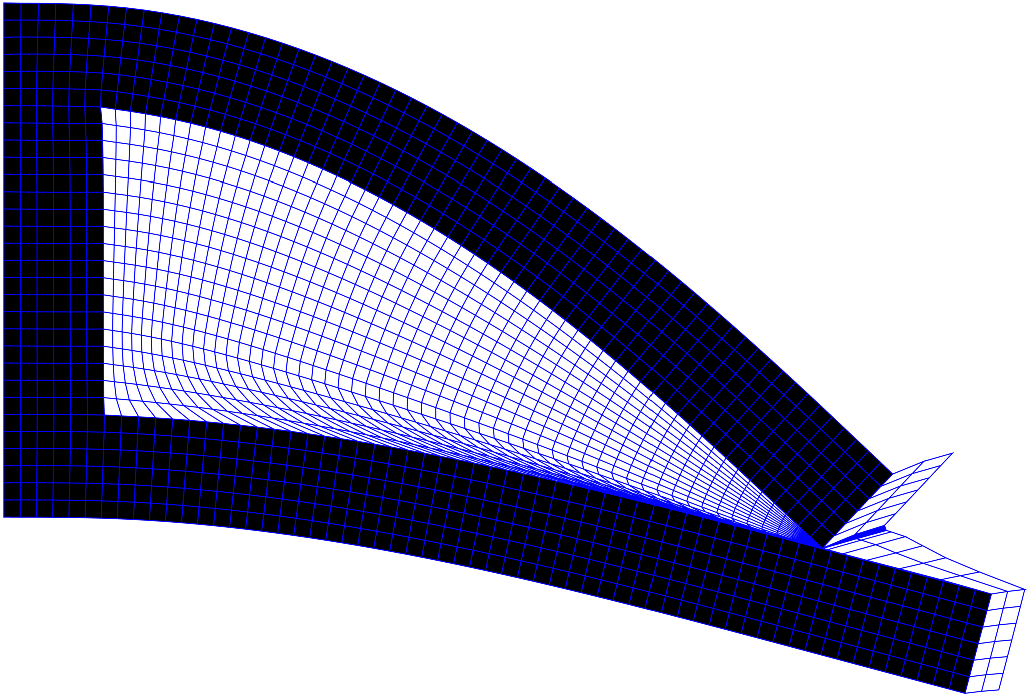}};
     \fill [pattern=north east lines] (-0.35,-0.35) rectangle (0,3.2);
     \draw [-] (0,-0.35) -- (0,3.2);
    \end{tikzpicture}
    \begin{tikzpicture}
     \node[anchor = north west, inner sep=0](img) at (0,2.75) {\includegraphics[clip,scale = 0.265]
     {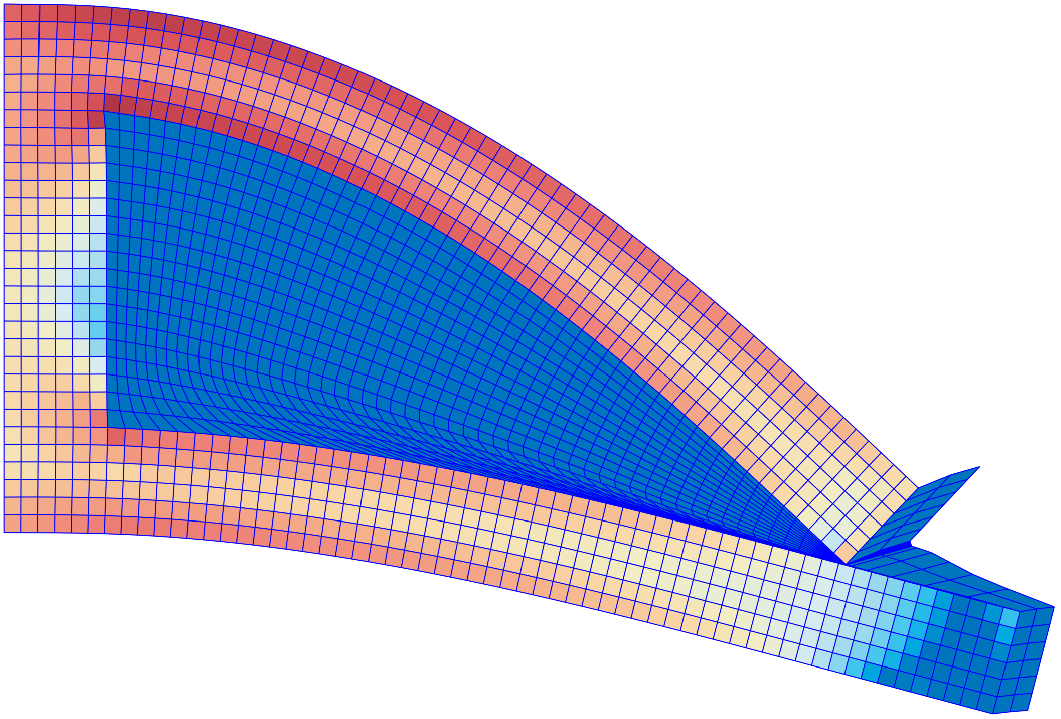}};
     \fill [pattern=north east lines] (-0.35,-0.35) rectangle (0,3.2);
     \draw [-] (0,-0.35) -- (0,3.2);
    \end{tikzpicture}
    \caption{$\Lambda = 1.0$}
    \label{fig:C_shape_c}
   \end{subfigure}
  \caption{\small{Deformed configurations (top row), and distribution of the SED (bottom row) for the C-shape example at three load steps. The SED is normalized with respect to the maximum domain value, and plotted in \emph{log}-scale}. Deep red corresponds to regions where the SED is highest, and deep blue to regions where the SED is lowest}
 \label{fig:CshapeResults1}
\end{figure*}

\begin{figure}[htbp]
 \centering
  \includegraphics[clip,scale = 0.475]
  {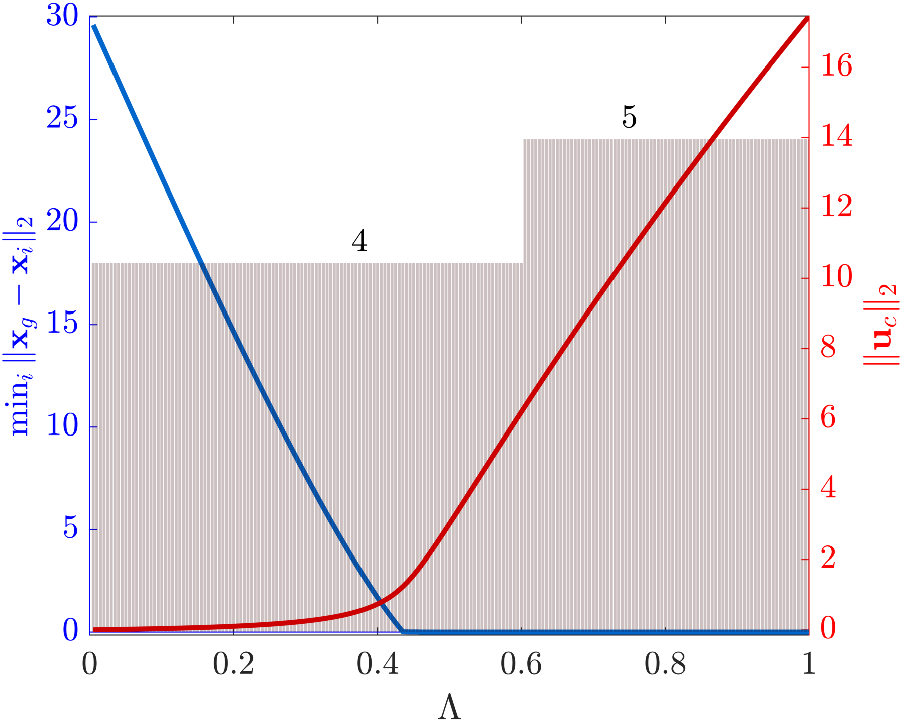}
 \caption{\small{Contact gap between the upper and lower beams of the C-shape configuration (blue curve plotted against left axis) and force transfer, linked to the displacement $\boldsymbol{u}_{c}$ (red curve plotted against right axis). The figure also shows the number of Newton iterations required to restore equilibrium, for each load step (shaded bar plot)}}
 \label{fig:CshapeResults2}
\end{figure}

The deformed configurations corresponding to three load levels are shown in \autoref{fig:CshapeResults1}, together with the domain distribution of the relative strain energy density (SED). \autoref{fig:CshapeResults2} shows the trend of the contact gap, defined as the minimum distance $\min_{i}\|\boldsymbol{x}_{g} - \boldsymbol{x}_{i}\|_{2}$ (see blue dot and line in \autoref{fig:exampleCshapeSetup}) and the norm of the absolute displacement of point $\boldsymbol{x}_{c}$ (see red dot in \autoref{fig:exampleCshapeSetup}). These plots show that there is no significant force transfer between the upper and lower beams until contact is engaged, at about $\Lambda\approx 0.4$. After this point the force transfer, and thus the displacement of the lower beam quickly increases. However, even for $\Lambda < 0.4$ the force transfer is not strictly zero and thus the lower beam still experiences some (very low) deformations due to the bending of the upper one. This is a fundamental feature of TMC when applied to optimization, as we will show in \autoref{sSec:topologyOptimizationExample}.

\autoref{fig:CshapeResults2} also displays the number of Newton iterations needed for achieving equilibrium at each load step (shaded bar plot). Four to five Newton iterations are needed through the whole loading process, showing the effectiveness of the \textsf{HuHu}-regularization even when contact and element deformation are very pronounced. We point out that, within our experimental setting we could achieve convergence also when using a slightly inconsistent tangent matrix, obtained by neglecting the last term in \eqref{eq:discretizedTangentStiffness}. This choice makes the tangent matrix symmetric, thus allowing for the use of more efficient assembly and solution methods \citep{ferrari-sigmund_20b_99LinesNewGeneration}. On the other hand, we observe a substantial increase in the number of Newton iterations required as the deformation grows, and higher sensitivity to parameters such as the load step-size (\texttt{nIncr}), and the \textsf{HuHu}-regularization weight (\texttt{alpha}).

The choice of $\alpha$ ensures that the TM negligibly affects the deformation of the solid, except when it becomes nearly fully compressed ($|\boldsymbol{F}| \approx 0$). Larger values of $\alpha$ generally improve the convergence of the nonlinear solver, but at the price of introducing errors in the physics. For a broader discussion about the influence of the material, load level, and regularization parameters on the convergence of this example we refer to \cite{faltus-etal_24a_tmcPneumaticallyActuated, frederiksen-etal_24a_topOptSelfContacting, wriggers-etal_25a_tmcFirstSecondElement, Thesis_Andreas_TMC}.

The provided code has been tested for the set of parameters in \autoref{tab:parametersCshape}, and we observed robust convergence also when applying moderate changes to the mesh resolution, material moduli and number of load increments. However, we caution that the implementation is by no means robust with respect to arbitrary changes of parameters, which may affect or even destroy the convergence behaviour. Therefore, we suggest users aiming for general use to couple the provided routines with more general and robust nonlinear solvers.

\section{Topology optimization framework}
 \label{Sec:TOframework}

Following density-based TO \citep{book:bendsoe-sigmund_2004}, we consider the design field $\rho(\boldsymbol{x})$ and the relative density field $\hat{\rho}(\boldsymbol{x})$, $\boldsymbol{x}\in \Omega:=\Omega_{s}\cup\Omega_{v}$. The two fields are linked by the linear PDE-based filter \citep{lazarov-sigmund_11a}, and the relaxed Heaviside projection \citep{wang-etal_11a_projectionMethodsRobust} 
\begin{align}
 \label{eq:filteringHeavisideProjection}
 \tilde{\rho}(\boldsymbol{x}) & =
 {\rm argmin}_{\varrho}
 \int_{\Omega}
 l^{2}_{\Omega}|\nabla\varrho|^{2} + (\rho-\varrho)^{2}
 \: {\rm d}\Omega \\
 \hat{\rho}(\boldsymbol{x}) & =
 \frac{\tanh(\beta\eta) + 
 \tanh(\beta(\tilde{\rho}(\boldsymbol{x})-\eta))}
 {\tanh(\beta\eta) + \tanh(\beta(1-\eta))} 
\end{align}
where $l_{\Omega} \geq 1$ is the filter radius (linked to the parameter \texttt{rmin} in the \textsf{Matlab} code), $\tilde{\rho}(\boldsymbol{x})$ is the intermediate field, and $\eta = [0,1]$, $\beta = [1, \infty)$ define the projection threshhold and curvature.

The relative density is used to parametrize the material properties, here through the RAMP interpolation \citep{stolpe-svanberg_01a}
\begin{equation}
 \label{eq:rampInterpolation}
  \gamma(\hat{\rho}(\boldsymbol{x})) = \gamma_{0} + (1-\gamma_{0})
  \frac{\hat{\rho}(\boldsymbol{x})}{1+p(1-\hat{\rho}(\boldsymbol{x}))}
\end{equation}
where $\gamma_{0}$ is the scaling of the material property over the void, and $p \geq 0$ is the penalization factor.

The interpolation \eqref{eq:rampInterpolation} enters the mechanical equilibrium through the SED
\begin{equation}
 \label{eq:interpolationSED}
  \tilde{W}(\hat{\rho},\boldsymbol{u}) = \gamma(\hat{\rho}) W(\boldsymbol{u}) + \frac{k_{r}}{2}
  \mathbb{H}\boldsymbol{u}\cdot\mathbb{H}\boldsymbol{u}
\end{equation}
such that the void region, representing the TM becomes significantly softer than the solid. To achieve this in the MATLAB code we interpolate both initial Lam\'{e} parameters according to \eqref{eq:rampInterpolation}, such that $\lambda = \lambda(\hat{\rho}(\boldsymbol{x}))$ and $\mu = \mu(\hat{\rho}(\boldsymbol{x}))$.

Also, we remark that the \textsf{HuHu} regularization term is not scaled by the material interpolation, and thus it depends on $\hat{\rho}$ only implicitly, through the displacement field $\boldsymbol{u} = \boldsymbol{u}(\hat{\rho})$.

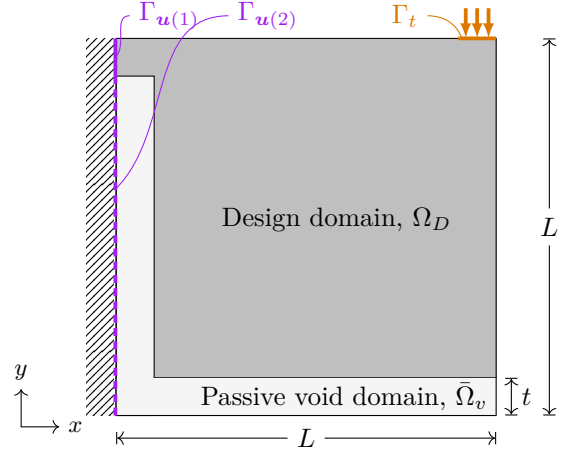
\begin{figure}[tb]
 \centering
  \begin{tikzpicture}
   \definecolor{customgray}{RGB}{245, 245, 245}
   \definecolor{custompurple}{RGB}{160, 32, 240}
   \definecolor{customorange}{RGB}{220, 120, 0}
   \fill [fill=gray!50] (0,0) rectangle (5,5);
   \draw [draw=black] (0,0) rectangle (5,0.5);
   \draw [fill=customgray, draw=black] (0,0) rectangle (.5,4.5);
   \fill [fill=customgray] (0,0) rectangle (5,0.5);
   \fill [pattern=north east lines] (-0.39,0.0) rectangle (-0.04,5.0);
   \draw [draw=black] (0,0) rectangle (5,5);
   \draw [color=custompurple, line width=0.5mm] (-0.01,4.5) -- (-0.01,5.0);
   \draw [color=custompurple, line width=0.5mm, densely dashed] (-0.01,0.0) -- (-0.01,4.5);
   \node (L) at (2.5,-0.3) {\normalsize $L$};
   \draw [|<-] (0,-0.3) -- (L);
   \draw [->|] (L) -- (5,-0.3);
   \node (H) at (5.7, 2.5) {\normalsize $L$};
   \draw [|<-] (5.7, 0) -- (H);
   \draw [->|] (H) -- (5.7,5);
   \node (t) at (5.4, 0.25) {\normalsize $t$};
   \draw [|<->|] (5.2, 0) -- (5.2,0.5);
   \node [color=black] (omega_s) at (2.9, 2.6) {Design domain, $\Omega_{D}$};
   \node (omega_v) at (3, 0.25) {Passive void domain, $\Omega_{v}$};
   \node [color=customorange] (gamma_t) at (3.8, 5.3) {\normalsize $\Gamma_{t}$};
   \draw [color=customorange] (gamma_t.east) to[out=0, in=110] (4.55,5);
   \draw [color=customorange, line width=0.5mm] (4.5,5) -- (5,5);
   \node [color=custompurple] (gamma_u1) at (0.7, 5.3) {\normalsize $\Gamma_{\boldsymbol{u}(1)}$};
   \node [color=custompurple] (gamma_u2) at (2.0, 5.3) {\normalsize $\Gamma_{\boldsymbol{u}(2)}$};
   \draw [color=custompurple] (gamma_u1.west) to[out=180, in=45] (-0.01,4.75);
   \draw [color=custompurple] (gamma_u2.west) to[out=180, in=45] (-0.01,3.0);
   \foreach \x in {4.6, 4.75, 4.9} {
   \draw [color=customorange, line width=0.5mm, -{Latex[length=2mm, width=1.5mm]}] (\x,5.4) -- (\x,5);
   }
  \end{tikzpicture}
 \caption{\small{Geometrical and mechanical setup for the minimum end-compliance TO example. The left edge $\Gamma_{\boldsymbol{u}(1)}\cup\Gamma_{\boldsymbol{u}(2)}$ is clamped, and we freeze the value $\rho(\boldsymbol{x}) = 0$ on the passive void region $\bar{\Omega}_{v}$}}
 \label{fig:exampleTopOpt}
\end{figure}

\subsection{End-compliance minimization with support contact}
 \label{sSec:topologyOptimizationExample}

We consider the task of minimizing the \textit{end-compliance} of a loaded structure subject to a volume constraint \citep{buhl-etal_00a_geometricNonlinearTO, kemmler-etal_05a_largeDeformationStability}. To promote convergence towards discrete designs, we impose the volume constraint on the \emph{dilated} field of control variables $\hat{\rho}^{(d)}$, computed by the relaxed Heaviside projection \eqref{eq:filteringHeavisideProjection} for a threshold value $\eta^{(d)} < \eta$. This does not necessarily amount to using a robust optimization approach, although it is inspired by it; we just further penalize intermediate densities by computing stiffness properties on a design which is slightly eroded (thus weaker) than the one used for computing the volume constraint \citep{frederiksen-etal_24a_topOptSelfContacting}.

The optimization problem reads as follows
\begin{equation}
 \label{eq:optFormulationMinEndCompliance}
  \begin{aligned}
    \min_{\rho} \quad & 
    c(\hat{\rho}) = 
    \Lambda_{\rm end} \ell(\boldsymbol{u}_{\rm end}) \\
    \text{s.t.} \quad & \mathcal{R}_{\rm int}(\hat{\rho}, \boldsymbol{u}_{\rm end};\delta \boldsymbol{u}) - \Lambda_{\rm end}\ell(\delta\boldsymbol{u}) = 0 \quad \forall\: \delta\boldsymbol{u} \\
    & v_{f}(\hat{\rho}^{(d)}) \leq \bar{v}_{f} \\
    & 0 \leq \rho(\boldsymbol{x}) \leq 1 \qquad \boldsymbol{x}\in\Omega
\end{aligned}
\end{equation}
where $\Lambda_{\rm end}$ is the end value of the load multiplier, $\ell(\boldsymbol{u})$ is defined in \eqref{eq:variationLF}, and $\bar{v}_{f}$ is the allowed maximum volume fraction.

To obtain a design with a sharper solid-void boundary we apply continuation to the Heaviside projection curvature parameter $\beta$, starting from $\beta=1$ and then gradually increasing it up to $\beta = 8$. Problem \eqref{eq:optFormulationMinEndCompliance} is solved by the Nested Analysis and Design approach \citep{book:haftka2012}. At each re-design step the sensitivity of the end-compliance to the variation of the control variable, say $\delta\hat{\rho}$ reads
\begin{equation}
 \label{eq:endComplianceSensitivityContinuous}
 \delta c(\delta\hat{\rho}) = 
 \mathcal{R}_{\rm int}(\hat{\rho}, \boldsymbol{u}_{\rm end};\delta\hat{\rho})\boldsymbol{\theta}(\hat{\rho}, \boldsymbol{u}_{\rm end})
\end{equation}
where $\boldsymbol{\theta}$ solves the adjoint system $\mathcal{K}_{T}(\hat{\rho},\boldsymbol{u}_{\rm end})\boldsymbol{\theta} = \Lambda_{\rm max} \ell(\boldsymbol{u})$, and $\mathcal{K}_{T}(\hat{\rho},\boldsymbol{u}_{\rm end})$ is the tangent stiffness operator evaluated for the end-displacement. The sensitivities w.r.t. the design field $\rho(\boldsymbol{x})$ are then obtained by applying the chain rule according to the relationships \eqref{eq:filteringHeavisideProjection}.

Upon discretization, these equations assume the well-known form (see, e.g., equations (13)-(14) in \cite{buhl-etal_00a_geometricNonlinearTO}). At each redesign step, the design field is updated through the simplified MMA-like approximation \citep{svanberg_87a}, adapted from \cite{ferrari-etal_21a_250LinesTopOptBuckling} and given in the routine \texttt{ocUpdate.m} (see \autoref{lst:ocUpdate}).

We consider the configuration in \autoref{fig:exampleTopOpt}. The square region is clamped along the whole left edge $\Gamma_{\boldsymbol{u}(1)}\cup \Gamma_{\boldsymbol{u}(2)}$, and loaded by a downward surface traction at the boundary $\Gamma_{t}$, with length $t = L/10$. With the end-magnitude set to $|q| = 40 \: {\rm N}$, we consider $\Lambda_{\rm end} = 1$. We stress that, even if solid material can connect to $\Gamma_{\boldsymbol{u}}$ only for the depth $t$, the whole left edge is clamped. This is fundamental when considering TMC modelling, as the optimizer will realize that the design can take advantage of the support stiffness through compression of the TM.

The optimizer updates the design variable field over the design domain $\Omega_{D}$, whereas we prescribe $\boldsymbol{\rho} = 0$ over the void region $\Omega_{v}$. However, small values of the intermediate ($\tilde{\boldsymbol{\rho}}$) and control fields ($\tilde{\boldsymbol{\rho}}$) propagate into $\Omega_{v}$ for a depth of $r_{\rm min}$, due to the PDE-filtering. This has the important effect of transferring a small force between the clamped edge and the design domain, informing the optimizer about the advantage of placing material near to the edge. On the other hand, at the right edge we add two rows of elements where all the design-related fields are set to zero ($\boldsymbol{\rho} = \tilde{\boldsymbol{\rho}} = \hat{\boldsymbol{\rho}} = 0$). This has been observed to stabilize further the convergence of the nonlinear solver, in line with the experience from the C-shape example.

\begin{table}[tb]
    \centering
    \caption{\small{Dimensions and parameters for the topology optimization example.}}
    \label{tab:topopt_parameters}
    \renewcommand*{\arraystretch}{1.3}
    \begin{tabular}{llll}
        \toprule
        Parameter & Code symbol(s) & Value(s) & Unit \\
        \midrule
        Domain length              & \texttt{L}              & $100$             & mm \\
        Void domain thickness      & \texttt{thk}            & $L/10$            & mm \\
        Young's modulus (solid)    & \texttt{E0}             & $100$             & MPa\\
        Poisson's ratio            & \texttt{nu}             & $0.3$             & -- \\
        TMC contrast               & \texttt{kv}             & $10^{-6}$         & -- \\
        RAMP penalization         & \texttt{qRamp}           & $4$              & --  \\
        \textsf{HuHu} parameter    & \texttt{alpha}          & $10^{-6}$         & -- \\
        Filter radius              & \texttt{rmin}           & $L/48$            & mm \\
        Proj. thresholds           & [\texttt{etaB, etaD}]   & $[0.5, 0.45]$     & -- \\
        Proj. sharpness            & \texttt{beta}           & $1 \rightarrow 8$ & -- \\
        Max. volume fraction       & \texttt{volfrac}        & $0.25$            & -- \\
        Number of elements         & [\texttt{nelx,nely}]    & $[162,160]$       & -- \\
        \bottomrule
    \end{tabular}
\end{table}

The user-defined parameters for replicating the following results by using \texttt{topTMC.m} are listed in \autoref{tab:topopt_parameters}.

\renewcommand\thesubfigure{\roman{subfigure}}
\begin{figure*}
 \centering
   \begin{subfigure}[b]{0.3\textwidth}
    \begin{tikzpicture}
    \node[anchor=north west, inner sep=0] (img) at (0,3.75)
    {\includegraphics[width = 3.76cm, trim={-0.025cm 0 0 0.05cm},clip]
    {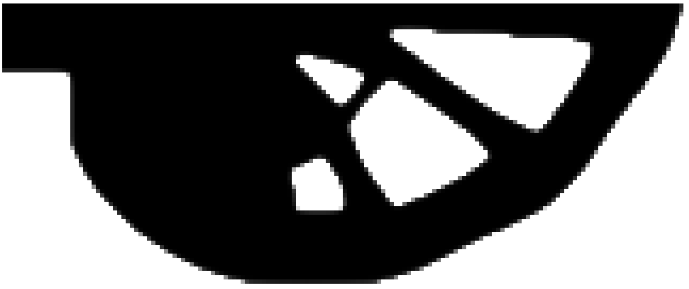}};
    \fill [pattern = north east lines] ( -0.375,-0.0) rectangle (0,3.75);
    \draw [draw=black] (0,0) rectangle (3.75, 3.75);
    \draw [-, line width=0.3mm] (0,-0.) -- (0, 3.75);
    \end{tikzpicture}
    \begin{tikzpicture}
    \node[anchor=north west, inner sep=0] (img) at (0,3.75)
    {\includegraphics[width = 3.76cm, trim={-0.025cm 0 0 0.05cm},clip]
    {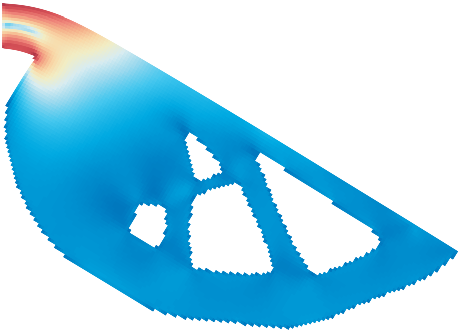}};
    \fill [pattern = north east lines] ( -0.375,-0.0) rectangle (0,3.75);
    \draw [draw=black] (0,0) rectangle (3.75, 3.75);
    \draw [-, line width=0.3mm] (0,-0.) -- (0, 3.75);
    \end{tikzpicture}
   \caption{\small{Linear elasticity $c = 210.12$}}
   \end{subfigure}
  \hfill
   \begin{subfigure}[b]{0.3\textwidth}
    \begin{tikzpicture}
    \node[anchor=north west, inner sep=0] (img) at (0,3.75)
    {\includegraphics[width = 3.76cm, trim={-0.025cm 0 0 0.05cm},clip]
    {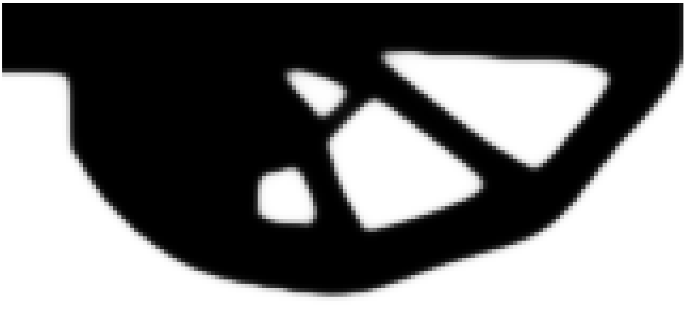}};
    \fill [pattern = north east lines] ( -0.375,3.35) rectangle (0,3.75);
    \draw [draw=black] (0,0) rectangle (3.75, 3.75);
    \draw [-, line width=0.3mm] (0,-0.) -- (0, 3.75);
    \end{tikzpicture}
    \begin{tikzpicture}
    \node[anchor=north west, inner sep=0] (img) at (-0.025,3.77)
    {\includegraphics[width = 3.56cm, trim={-0.025cm 0 0 0.05cm},clip]
    {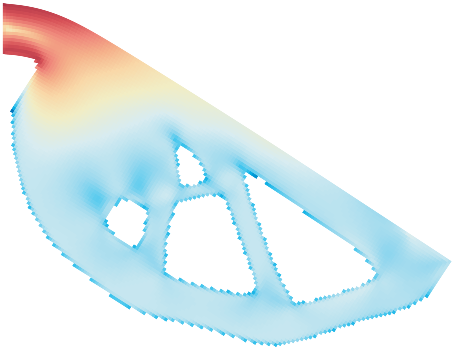}};
    \fill [pattern = north east lines] ( -0.375,3.35) rectangle (0,3.75);
    \draw [draw=black] (0,0) rectangle (3.75, 3.75);
    \draw [-, line width=0.3mm] (0,-0.) -- (0, 3.75);
    \end{tikzpicture}
    \caption{\small{Nonlinear elasticity $c = 191.96$}}
    \end{subfigure}
   \hfill
   \begin{subfigure}[b]{0.3\textwidth}
    \begin{tikzpicture}
    \node[anchor=north west, inner sep=0] (img) at (0,3.76)
    {\includegraphics[width = 3.76cm, trim={-0.025cm 0 0 0.05cm},clip]
    {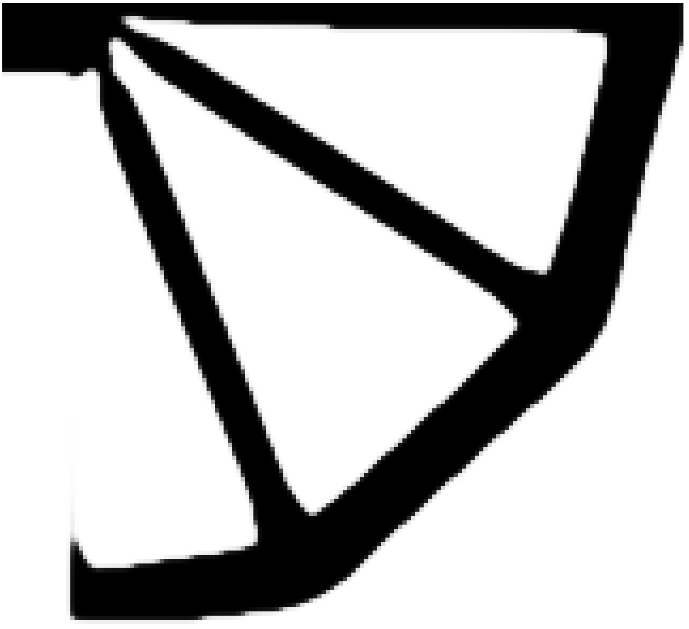}};
    \fill [pattern = north east lines] ( -0.375,-0.0) rectangle (0,3.75);
    \draw [draw=black] (0,0) rectangle (3.75, 3.75);
    \draw [-, line width=0.3mm] (0,-0.) -- (0, 3.75);
    \end{tikzpicture}
    \begin{tikzpicture}
    \node[anchor=north west, inner sep=0] (img) at (-0.005,3.76)
    {\includegraphics[width = 3.72cm, trim={-0.025cm 0 0 0.05cm},clip]
    {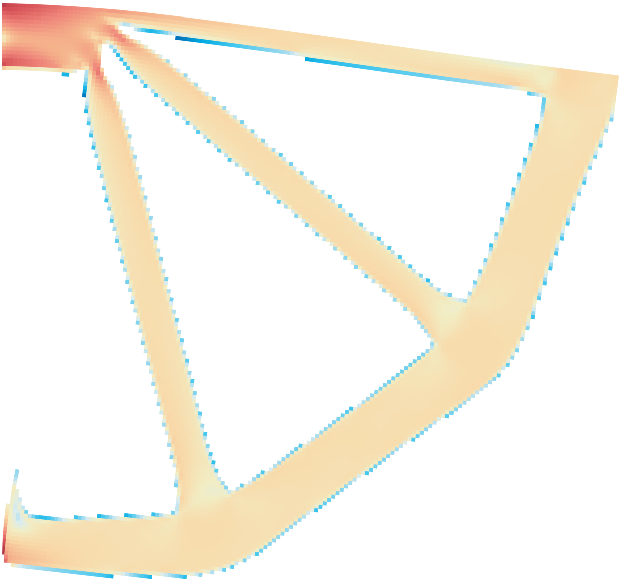}};
    \fill [pattern = north east lines] ( -0.375,-0.0) rectangle (0,3.75);
    \draw [draw=black] (0,0) rectangle (3.75, 3.75);
    \draw [-, line width=0.3mm] (0,-0.) -- (0, 3.75);
    \end{tikzpicture}
    \caption{\small{Contact modelling $c = 43.70$}}
   \end{subfigure}
    \caption{\small{Results for the TO example solved with three modelling assumptions. Contact is introduced, by the TMC, only in (iii), and the design clearly takes advantage of contact with the left edge support. The end-compliance value is shown in the subcaptions, and the SED (bottom row) is normalized with respect to the maximum domain value, and plotted in $log$-scale. Deep red corresponds to highest SED and deep blue to lowest SED}}
 \label{fig:TopOptResults1}
\end{figure*}
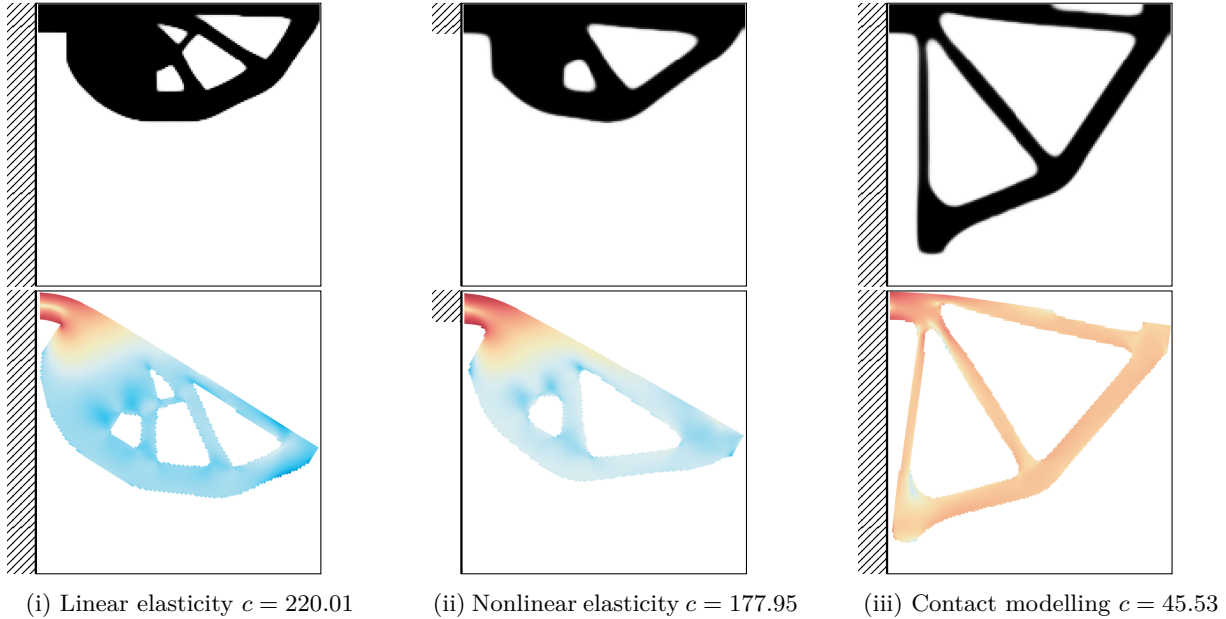

We solve the example using three different modelling assumptions:
\begin{itemize}
 \item[(i)] first, we mimic the solution corresponding to linearized elasticity by setting a very small end load factor: $\Lambda_{\rm end} = 0.01$. All qualitative results can then be  obtained by multiplying the end-force and end-displacement by $\Lambda^{-1}_{\rm end}$;
 \item[(ii)] then, we extend to nonlinear elasticity, using the neo-Hookean material law \eqref{eq:sedNeoHookean}, but without contact modelling. This is achieved by simply removing the clamping on the portion $\Gamma_{\boldsymbol{u}(2)}$ at the left edge. We stress that, in this case we must keep the $\textsf{HuHu}$ stabilization term, to avoid convergence issues due to highly distorted elements which may appear over regions with intermediate densities, during the optimization \citep{wang-etal_14a_interpolationLargeStrain};
 \item[(iii)] finally, we add contact modelling by restoring clamping over the whole region $\Gamma_{\boldsymbol{u}(1)}\cup\Gamma_{\boldsymbol{u}(2)}$.
\end{itemize}

Results for these three cases are collected in \autoref{fig:TopOptResults1}. For both cases (i) and (ii), the optimizer has no information about the stiffening contribution given by the portion $\Gamma_{\boldsymbol{u}(2)}$ of the supported edge. The only effective force transfer occurs over $\Gamma_{\boldsymbol{u}(1)}$, which is directly connected to the design domain. Thus, solid material is placed only in a shallow region close to the top edge of $\Omega_{D}$ and we see that the structure shows large localized deformations at the clamped region.

\begin{figure}[tb]
 \centering
 \includegraphics[width=\linewidth]
 {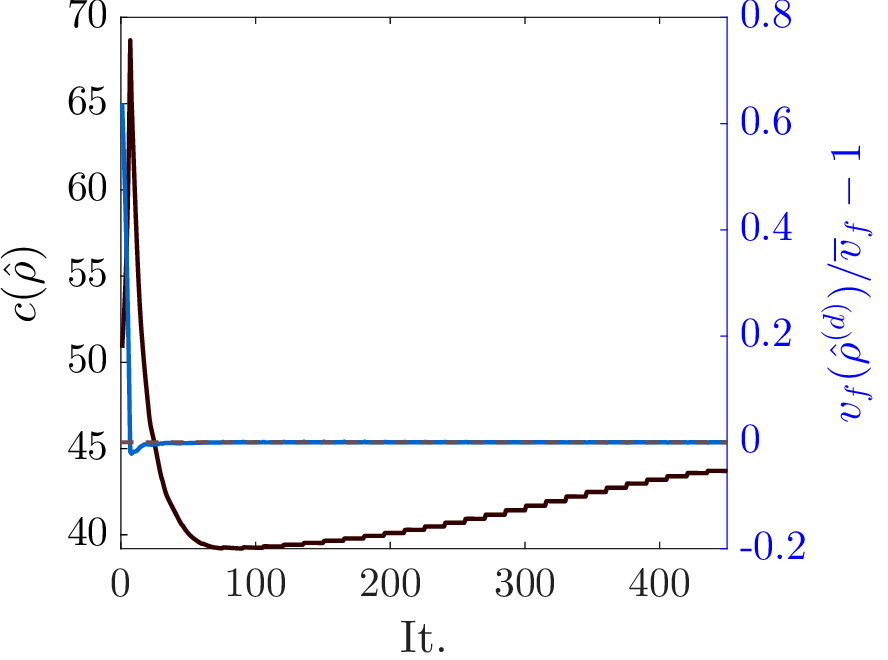}
 \caption{\small{Optimization history corresponding to the case of \autoref{fig:exampleTopOpt} (iii). The end-compliance and the volume constraint are plotted against the left and right axis, respectively}}
 \label{fig:exampleTopOpt_convergence}
\end{figure}

For the present setup, considering large deformations and hyperelastic response gives only minor changes in the resulting structural configuration, but a 10\% improvement in the optimized end-compliance. The \textsf{HuHu}-regularization is still very beneficial for this case, avoiding ill-conditioning of the equilibrium system due to the huge mesh distortion happening near the top left corner.

On the other hand, when contact is introduced by the TMC modelling (iii), the optimizer utilizes the advantage of placing material in contact with the left wall, also over the region $\Gamma_{\boldsymbol{u}(2)}$. Therefore, a deeper design develops to engage contact at an earlier stage, reducing the deformation near the top-left corner and at the loaded edge.

For case (iii), the evolution of the objective and constraint functions in the optimization progresses are shown in \autoref{fig:exampleTopOpt_convergence}. The force transfer happens through the upper clamped region and the contact point at the bottom. As a result, the SED is almost evenly distributed over the final design, and the compliance is reduced to nearly $22\%$ of that of the nonlinear design in case (ii), where contact was not considered. Interestingly, a fourth vertical bar, connecting the upper-left solid region with the lower left in contact also develops in the early optimization stages. Then, it slowly disappears as $\beta$ increases, as the lower part is sufficient to take advantage of the contact region.

\begin{figure*}[t]
 \centering
  \begin{subfigure}[t]{0.325\linewidth}
  \begin{tikzpicture}
   \definecolor{customgray}{RGB}{245, 245, 245}
   \definecolor{custompurple}{RGB}{160, 32, 240}
   \definecolor{customorange}{RGB}{220, 120, 0}
   \draw [draw=black] (0,0) rectangle (5,5);
   \fill [fill=gray!50] (0,0) rectangle (5,5);
   \filldraw[fill=customgray, opacity=0.5, draw=black]
   (0.0,2.25) --
   (5.0,2.25) --
   (5.0,2.75) --
   (3.5,2.75) --
   (3.5,4.50) --
   (0.5,4.50) --
   (0.0,4.50) -- cycle;
   \filldraw[fill=black, opacity=1.0, draw=black]
   (0.0,5.00) --
   (5.0,5.00) --
   (5.0,4.75) --
   (0.0,4.75) -- cycle;
   \fill [pattern=north east lines] (-0.25,0.0) rectangle (-0.05,5.0);
   \draw [color=custompurple, line width=0.5mm] (-0.01,0.0) -- (-0.01,5.0);
   \node (L) at (2.5,-0.5) {\normalsize $L$};
   \draw [|<-] (0,-0.5) -- (L);
   \draw [->|] (L) -- (5,-0.5);
   \node (t) at (5.4, 4.75) {\normalsize $t$};
   \draw [|<->|] (5.2, 4.5) -- (5.2,5.0);
   \node (t) at (5.4, 2.5) {\normalsize $t$};
   \draw [|<->|] (5.2, 2.25) -- (5.2,2.75);
   \node [color=black] (omega_s) at (2.9, 1.6) {Design domain, $\Omega_{D}$};
   \node (omega_v) at (3, 2.5) {Passive void domain, $\Omega_{v}$};
   \node [color=black] (psd) at (2.5, 6.0) {Passive solid domain ($\boldsymbol{\rho} = 1$)};
   \draw [color=black] (psd.south) to[out=0, in=125] (2.50,5.05);
   \node [color=customorange] (gamma_t) at (3.8, 5.3) {\normalsize $\Gamma_{t}$};
   \draw [color=customorange] (gamma_t.east) to[out=0, in=110] (4.55,5);
   \draw [color=customorange, line width=0.5mm] (4.5,5) -- (5,5);
   \node [color=custompurple] (gamma_u1) at (0.7, 5.3) {\normalsize $\Gamma_{\boldsymbol{u}}$};
   \draw [color=custompurple] (gamma_u1.west) to[out=180, in=45] (-0.01,4.75);
   \foreach \x in {4.6, 4.75, 4.9} {
   \draw [color=customorange, line width=0.5mm, -{Latex[length=2mm, width=1.5mm]}] (\x,5.4) -- (\x,5);}
  \end{tikzpicture}
 \end{subfigure}
 \hspace*{\fill}
 \begin{subfigure}[t]{0.325\linewidth}
  \begin{tikzpicture}
  \node[anchor=north west, inner sep=0] (img) at (0,5.0)
   {\includegraphics[width = 5.0cm, trim={-0.025cm 0 0 0.05cm},clip]
  {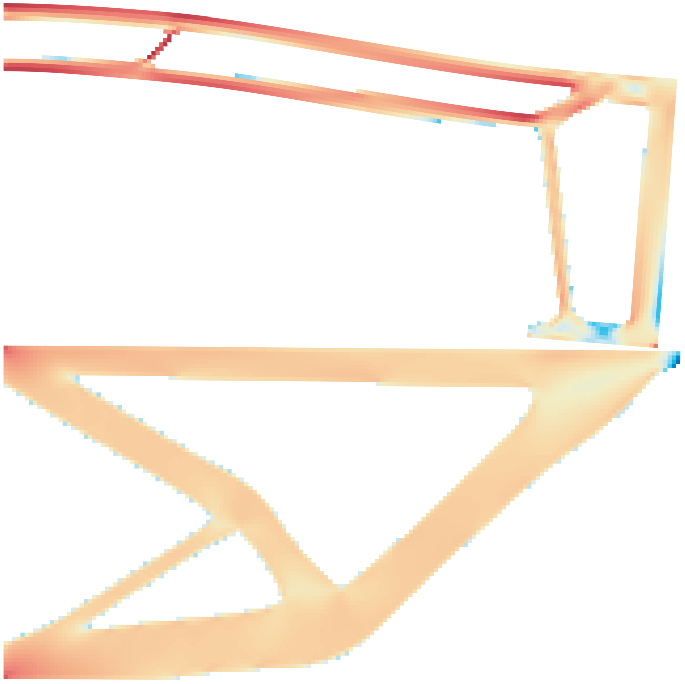}};
  \fill [pattern = north east lines] (-0.25, 0.0) rectangle (0, 5.0);
  \draw [draw=black] (0,0.0) rectangle (5.0,5.0);
  \node (L) at (2.5,-0.5) {\normalsize $c = 9.246$};
  \end{tikzpicture}
 \end{subfigure}
 \hspace*{\fill}
 \begin{subfigure}[t]{0.250\linewidth}
  \begin{tikzpicture}
  \node[anchor=north west, inner sep=0] (img) at (0,5.0)
   {\includegraphics[width = 4.1cm, trim={0 0 0 0},clip]
  {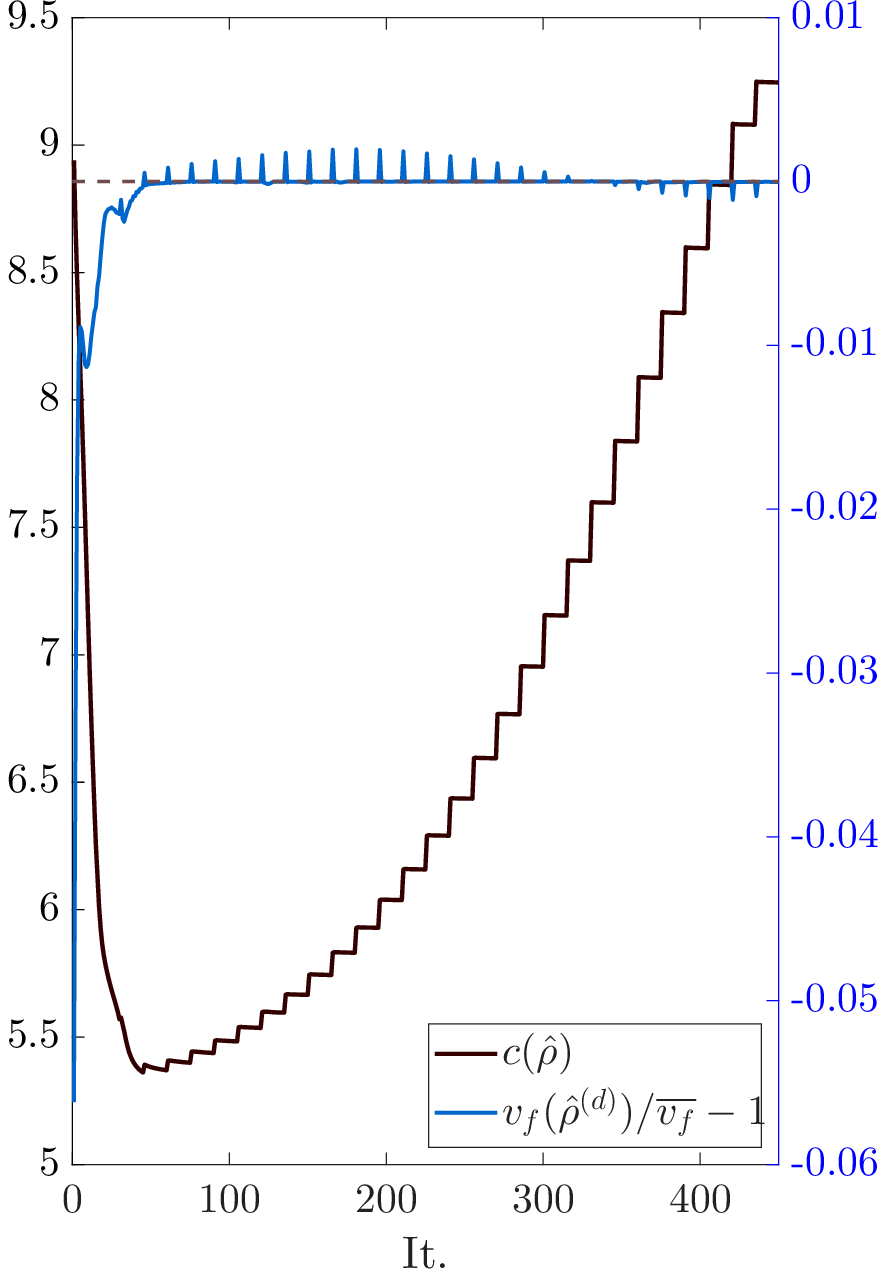}};
  \end{tikzpicture}
 \end{subfigure}
 \caption{\small{Setup for the minimum end-compliance TO example modified to achieve self-contact (left). Optimized design, deformed and with the contour plot of the relative SED corresponding to the end load magnitude (middle). Deep red corresponds to highest SED and deep blue to lowest SED. Convergence of the objective and constraint functions (right)}}
 \label{fig:exampleTopOpt_selfcontact}
\end{figure*}

\subsubsection{Example with self-contact}
 \label{ssSec:selfContacExample}

Finally, we modify the previous example such that the optimized design will exhibit self-contact. We refer to the configuration in \autoref{fig:exampleTopOpt_selfcontact} where, compared to that of \autoref{fig:exampleTopOpt}, the passive void region $\Omega_{v}$ now splits the design domain into two disconnected parts. Also, a layer of passive solid elements, where $\boldsymbol{\rho} = 1$, is prescribed for a depth of $t/2$ at the upper edge. The whole left edge is clamped and the load is applied at the same location as before. These changes are obtained by replacing a few lines in the main code in \autoref{lst:topTMC}, with those collected in \autoref{lst:cshapeTMC} (the specific lines are reported in the green comments by "(\texttt{L\#})").

All the other parameters are kept as in the previous example.

\begin{lstlisting}[basicstyle=\footnotesize\ttfamily,breaklines=true,numbers=none,frame=single, 
caption={Changes to \autoref{lst:topTMC} needed for running the self-contact example}
\label{lst:cshapeTMC}]
% (L4) # of elements in (x,y) directions
[nelx,nely] = deal(162,162);
% (L29) set of passive solid elements
pasS = feP.elNrs(1:thk/4,:);
% (L30-32) set of passive void elements
pasV=unique(union(feP.elNrs(thk+1:nely/2,1:8*thk), ...
feP.elNrs( nely/2-thk:nely/2,:)));
pasVr = unique(union(feP.elNrs(:,end-1:end),feP.elNrs(end-1:end,:)));
% (L37) apply external tractions
feP.F0(ld,1) = -1e-2*E0/(length(ld)-1);
% (L64) intialize design variables
x = zeros(feP.nEl,1); x(pasS) = 1; x(act)=0.35;
\end{lstlisting}

The optimized design and its deformation corresponding to the end load are displayed in the middle and right plots of \autoref{fig:exampleTopOpt_selfcontact}. 

A design which is intuitive for a clamped beam supporting a tip load develops in the lower part of the design domain, providing stiffness to the pillar developing in the upper part, when it comes to contact. The optimizer clearly prefers to utilize material for this lower supporting beam rather than on a more intricate design in the upper part. This is expected, since the only effective load transfer mechanism in the upper part comes from the leftward supported region, which has limited depth of $t$.

This example, which is considerably more complex compared to the previous one, shows the key role of TMC modelling for transferring information between the disjointed parts of the design domain, to achieve a design showing self-contact.

\section{Discussion and conclusions}
 \label{Sec:disussionConclusions}

We have presented the Third Medium Contact (TMC) model built into a \textsf{Matlab} code for mechanical analysis and topology optimization of geometrically nonlinear, hyperelastic structures. 

The self-contained implementation shows that the TMC method, together with the \textsf{HuHu}-regularization of void regions, can be easily integrated in density-based topology optimization. These methods are demonstrated on two examples, which are readily reproducible with the provided code, involving analysis and topology design of structures exhibiting contact.

The implementation prioritizes clarity and educational value over generality and computational efficiency. However, several extensions are possible with minor changes to the code. For instance:
\begin{itemize}
 \item Switching to other material laws \citep{klarbring-stromberg_13a_hyperelasticTO} is easily achieved by changing the Piola stress and tangent moduli expressions in the routine \texttt{assembleKtFi.m};
 \item Using higher-order finite elements, and the associated \textsf{HuHu-LuLu}-regularization, requires rather intuitive changes to the routines \texttt{initializeFEA.m} and \texttt{assembleKtFi.m};
 \item General and robust path-following methods and nonlinear solvers can be introduced in the \texttt{solveIncrIter.m} routine, keeping the tangent operator and residual vector computed by \texttt{assembleKtFi.m};
 \item The extension to 3D problem, and use of efficient iterative solvers \citep{amir-etal_14a_multigridCGTO} should also be straightforward.
\end{itemize}

We envision that extensions to more advanced applications or physics, such as frictional contact \citep{frederiksen-etal_24a_addingFriction} or multi-physics \citep{dalklint-etal_25a_topOptThermoRegulators} is also possible, albeit with heavier changes to the base code provided here.

\backmatter

\bmhead{Acknowledgments}
This research was supported by nTopology Inc. and Independent Research Fund Denmark through the TopCon Project (case number 1032-00228B). The authors also acknowledge the support from the Villum Foundation, through the Villum Investigator Project ``AMSTRAD'' (VIL54487).

\bmhead{Funding information}
nTopology Inc., Independent Research Fund Denmark  ``TopCon Project'' (case number 1032-00228B), Villum Investigator Project ``AMSTRAD'' (VIL54487).

\bmhead{Author contributions}
All authors have contributed to the work equally.

\bmhead{Replication of results}
The results presented can be replicated with the \textsf{Matlab} codes listed in the paper, 

\bmhead{Data availability}
The \textsf{Matlab} codes can be downloaded at \url{https://www.topopt.mek.dtu.dk/apps-and-software}.

\bmhead{Conflict of interest}
We have no conflict of interest to disclose.


\onecolumn

\begin{appendices}

\section{Overview of the \textsf{Matlab} implementation}
 \label{Sec:matlabImplementation}

The code presented includes the following main scripts and functions:
\begin{table}[h!]
    \renewcommand{\arraystretch}{1.2}
    \setlength{\tabcolsep}{8pt} 
    \begin{tabular}{p{.3\linewidth}p{.6\linewidth}}
        \texttt{cshapeTMC.m}     & Script solving the C-shape analysis example of \autoref{sSec:analysisCshape} \\
        \texttt{topTMC.m}        & Script solving the TO example of \autoref{sSec:topologyOptimizationExample} \\
        \texttt{initializeFEA.m} & Function setting up the discretization and FE operators \\
        \texttt{assembleKtFi.m}  & Function assembling the tangent stiffness matrix and internal force vector \\
        \texttt{solveIncrIter.m} & Function solving the incremental nonlinear equilibrium problem \\
        \texttt{ocUpdate.m}      & Function for the MMA-like design update (cfr. \cite{ferrari-etal_21a_250LinesTopOptBuckling}) \\
        \texttt{plotDeformed.m}  & Function plotting the current configuration
    \end{tabular}
   \vspace{-18pt}
\end{table}

In the following we list the \textsf{Matlab} code for each of the above scripts and functions, followed by a detailed explanation of the operations performed. For the sake of brevity, the codes listed only contain the essential ``inline'' comments. The same files, with more exhaustive comments, can be downloaded from the TopOpt webpage.

 \begin{lstlisting}[basicstyle=\footnotesize\ttfamily,breaklines=true,numbers=left,frame=single, 
caption={Main script for running the contact analysis on the C-shape example}
\label{lst:cshapeTMC}]
clear; close all; clc; format long;
% USER-DEFINED DATA AND ANALYSIS PARAMETERS >>>>>>>>>>>>>>>>>>>>>>>>>>>>>>>
% Domain size and discretization ------------------------------------------
[ Lx, Ly ] = deal( 100, 50 );      % domain lengths in the (x,y) directions
[ nelx, nely ] = deal( 62, 30 );        % # of elements in (x,y) directions
% Material properties (Young's modulus, contrast, Poisson ratio, HuHu coef)
[ E0, kv, nu, alpha ] = deal( 100, 1e-6, 0.3, 1e-6 );
% Set the nonlinear solver parameters -------------------------------------
nlP.lambdaMax = 1.0;                    % end value of the load multiplier
nlP.nIncr     = 100;                 % number of (constant) load increments
nlP.maxIter   =  25;           % maximum # of Newton equilibrium iterations
nlP.tolRelRes = 1e-8;          % convergence tolerance on relative residual
% PERFORM PRELIMINARY COMPUTATIONS AND PROBLEM SETUP >>>>>>>>>>>>>>>>>>>>>>
% Compute discretization and all displacement-independent FE operators ----
feP = initializeFEA( Lx, Ly, nelx, nely, E0, nu, alpha );
% Domain definition (specific for the the C-shape example) ----------------
thk  = nely / 5;            % thickness of the C-frame walls (see Figure 3)
xPhys = ones( nely, nelx ) .* kv;       % initialize TM material everywhere
xPhys( [ 1 : thk, end - thk + 1 : end ], 1 : 2 * nely ) = 1; % solid part 1
xPhys( :, 1 : thk ) = 1;                                     % solid part 2
% Boundary conditions and loads (specific for the the C-shape example) ----
fix  = reshape( 2 * feP.ndNrs( :, 1 ) - [ 1, 0 ], [], 1 );% clamp left edge
feP.free = setdiff( 1 : feP.nDof, fix );                 % set of free DOFs
ld = 2 * feP.ndNrs( 1, end - thk - 1 : end - mod(nelx,nely) );% loaded DOFs
feP.F0 = zeros( feP.nDof, 1 );
feP.F0( ld, 1 ) = -3e-2*E0/( length( ld ) - 1 ); % apply external tractions
[ feP.F0( ld( 1 ) ), feP.F0( ld( end ) ) ] = deal( feP.F0( ld( 1 ) ) / 2 );
% PERFORM NONLINEAR ANALYSIS >>>>>>>>>>>>>>>>>>>>>>>>>>>>>>>>>>>>>>>>>>>>>>
U = solveIncrIter( xPhys, xPhys, nlP, feP );
\end{lstlisting}

 \begin{lstlisting}[basicstyle=\footnotesize\ttfamily,breaklines=true,numbers=left,frame=single,caption={Main script for running topology optimization with TMC contact modeling}\label{lst:topTMC}]
clear; close all; clc;
% USER-DEFINED DATA AND ANALYSIS PARAMETERS >>>>>>>>>>>>>>>>>>>>>>>>>>>>>>>
% Domain size and discretization ------------------------------------------
[ Lx, Ly ] = deal( 100, 100 );     % domain lengths in the (x,y) directions
[ nelx, nely ] = deal( 162, 160 );      % # of elements in (x,y) directions
% Material properties (Young's modulus, contrast, Poisson ratio, HuHu coef)
[ E0, Emin, nu, alpha ] = deal( 100, 1e-6, 0.3, 1e-6 );
% Set the nonlinear solver parameters -------------------------------------
nlP.lambdaMax = 1.0;                     % end value of the load multiplier
nlP.nIncr = 200;                     % number of (constant) load increments
nlP.maxIter = 50;              % maximum # of Newton equilibrium iterations
nlP.tolRelRes = 1e-8;          % convergence tolerance on relative residual
% Density-based TopOpt parameters -----------------------------------------
volfrac = 0.25;                           % maximum allowed volume fraction
rmin = nelx / 48;                 % minimum filter radius (physical length) 
qRamp = 4;               % initial penalty value for the RAMP interpolation
[ etaB, etaD, beta ] = deal( 0.50, 0.45, 1 );  % Heaviside proj. parameters
maxit = 500;                             % maximum number of redesign steps
convTO = 1e-3;       % convergence tolerance on max design-variables change
ocPar = [ 0.05, 0.7, 1.2 ];      % parameters for the MMA-like design update
% beta-continuation scheme {val_max, val_multiplier, steps_per_fix_val} ---
betaCnt = { 15, 1.075, 8 };
% Compute discretization and all displacement-independent FE operators ----
feP = initializeFEA( Lx, Ly, nelx, nely, E0, nu, alpha );
% Define design domain, loads, boundary conditions and passive regions ----
thk = nely / 10;
fix = reshape( 2 * feP.ndNrs( :, 1 ) - [ 1, 0 ], [], 1 ); % restrained DOFs
feP.free = setdiff( 1 : feP.nDof, fix );                 % set of free DOFs
pasS = [];                                  % set of passive solid elements
pasV = unique( union( feP.elNrs( thk + 1 : end, 1 : thk ), ...
    feP.elNrs( end - thk : end, : ) ) );     % set of passive void elements
pasVr = feP.elNrs( :, end - 1 : end );   % fixed void elements at the right
pasEl = [ pasS( : ); pasV( : ); pasVr( : ) ];         % whole set of passive elements
act = setdiff( ( 1 : feP.nEl )' , pasEl );         % set of active elements
ld = 2 * feP.ndNrs( 1, end - thk - 2 : end - 2 );      % set of loaded DOFs
feP.F0 = zeros( feP.nDof, 1 );        % initialize vector of external loads
feP.F0( ld, 1 ) = -4e-2*E0/( length( ld ) - 1 ); % apply external tractions
[ feP.F0( ld( 1 ) ), feP.F0( ld( end ) ) ] = deal( feP.F0( ld( 1 ) ) / 2 );
% Setup of density filter -------------------------------------------------
L0  = rmin / 2 / sqrt( 3 );
kef = L0^2*[4,-1,-2,-1;-1,4,-1,-2;-2,-1,4,-1;-1,-2,-1,4]/6+...
    [4,2,1,2;2,4,2,1;1,2,4,2;2,1,2,4]/36;
cVec = reshape( feP.ndNrs( 1 : end - 1, 1 : end - 1 ), feP.nEl, 1 );
cMat = repmat( cVec, 1, 4 ) + repmat( [ 0, nely + [ 1 : 2 ], 1 ], feP.nEl, 1 );
iKf = reshape( kron( cMat, ones( 4, 1 ) )', 16 * feP.nEl, 1 );
jKf = reshape( kron( cMat, ones( 1, 4 ) )', 16 * feP.nEl, 1 );
sKf = reshape( kef( : ) * ones( 1, feP.nEl ), 16 * feP.nEl, 1 );
Af  = sparse( iKf, jKf, sKf );
dAf = decomposition( Af, 'chol', 'lower' );
iTF = reshape( cMat, 4 * feP.nEl, 1 );
jTF = reshape( repmat( [ 1 : feP.nEl ], 4, 1 )', 4 * feP.nEl, 1 );
sTF = repmat( 1 / 4, 4 * feP.nEl, 1 );
Tf  = sparse( iTF, jTF, sTF );
% Inline functions for filtering, projection, interpolation & continuation
applyF = @( v ) Tf'*(dAf\(Tf*v(:)));            % apply PDE-based filtering
prj = @(v,eta,beta) (tanh(beta*eta)+tanh(beta*(v(:)-eta)))./...
    (tanh(beta*eta)+tanh(beta*(1-eta)));     % relaxed Heaviside projection
dprj = @(v,eta,beta) beta*(1-tanh(beta*(v-eta)).^2)./...
    (tanh(beta*eta)+tanh(beta*(1-eta))); % relaxed Heaviside proj. x-derivative
RAMP = @(v) (Emin+(1-Emin).*v./(1+qRamp.*(1-v)));      % RAMP interpolation
dRAMP = @(v) (1-Emin)*(qRamp+1)./(-1+(v-1)*qRamp).^2; % RAMP interpolation x-derivative
cnt = @(v,vCnt,l) min(v*(mod(l,vCnt{1})~=0)+v*(mod(l,vCnt{1})==0)*vCnt{2},vCnt{3});
% Initialize design variables and other arrays ----------------------------
x = zeros( feP.nEl, 1 ); x( pasS ) = 1; x( act ) = 0.5;
dv = ones( feP.nEl, 1 ) ./ ( volfrac * feP.nEl ); dv( pasEl ) = 0;
Uadj = zeros( feP.nDof, 1 ); chDVs = 1; loop = 0;
% PERFORM NONLINEAR ANALYSIS FOR THE INITIAL DESIGN >>>>>>>>>>>>>>>>>>>>>>>
xTilde = applyF( x ); xTilde( pasVr ) = 0;  % filter design variables (DVs)
xPhys  = prj( xTilde, etaB, beta );    % project the filtered variables
U = solveIncrIter( xPhys, RAMP( xPhys ), nlP, feP );
% START TopOpt REDESIGN PROCESS >>>>>>>>>>>>>>>>>>>>>>>>>>>>>>>>>>>>>>>>>>>
while ( chDVs > convTO && loop < maxit )
    loop   = loop + 1;                           % update iteration counter
    xTilde = applyF( x ); xTilde( pasVr ) = 0;                 % filter DVs
    xPhys  = prj( xTilde, etaB, beta );    % project the filtered variables
    % Solve nonlinear FE problem for the current design -------------------
    nlP.nIncr = 1;     % only solve for the final load step (end magnitude)
    U = solveIncrIter( xPhys, RAMP( xPhys ), nlP, feP, U );
    % Compute objective and constraint sensitivities ----------------------
    % ------------------------------- assemble and solve the adjoint system
    [ Kt, dFint ] = assembleKtFi( feP, U( feP.cDofMat )', RAMP( xPhys ), 'ad', dRAMP( xPhys ) );
    Uadj( feP.free, :) = - Kt( feP.free, feP.free ) \ ( nlP.lambdaMax * feP.F0( feP.free, 1 ) );
    % ------------------------------- compute and back-filter sensitivities
    dc  = ( Uadj( feP.free )' * dFint( feP.free, : ) )';
    dg0 = applyF( dc( : ) .* dprj( xTilde, etaB, beta ) );
    dg1 = applyF( dv( : ) .* dprj( xTilde, etaD, beta ) );
    % Update design variables (DVs) by the symplified MMA -----------------
    g0 = nlP.lambdaMax * feP.F0' * U( : );
    g1 = mean( prj( xTilde, etaD, beta ) ) / volfrac - 1;
    if loop == 1, xO = x( act ); xO1 = xO; asLU = []; low = []; upp = []; end
    [ xMMA, asLU, lmid ] = ocUpdate( loop, x( act ), dg0( act, 1 ), g1, dg1( act, : ), ...
        ocPar, xO, xO1, asLU, beta, 0 );
    xO1 = xO; xO = x( act ); x( act ) = xMMA;
    chDVs = max( abs( xMMA - xO ) );
    fprintf( '\n\n TopOpt It.:%1i g0:%0.3f g1:%0.3f ch.:%0.2e qRAMP:%1i beta:%0.3f\n', ...
        loop, g0, g1, chDVs, qRamp, beta ); 
    % Apply continuation to the beta projection parameter -----------------
    beta = cnt( beta, betaCnt, loop ); g0Vec( loop, 1 ) = g0; g1Vec( loop, 1 ) = g1;
end
\end{lstlisting}

 \begin{lstlisting}[basicstyle=\footnotesize\ttfamily,breaklines=true,numbers=left,frame=single, 
caption={OC-based update, based on a simplified MMA-like expansion of the objective and constraint}
\label{lst:ocUpdate}]
function [ x, as, lmid ] = ocUpdate( loop, xT, dg0, g1, dg1, ocPar, xOld, xOld1, as, beta, restartAs )
% Definition of the asymptotes and move limits ----------------------------
[ xU, xL ] = deal( min( xT + ocPar( 1 ), 1 ), max( xT - ocPar( 1 ), 0 ) );
if (loop < 2.5 || restartAs == 1 )
    asyLU = xT + [ -0.5, 0.5 ] .* ( xU - xL ) ./ ( beta + 1 );
else
    tmp = ( xT - xOld ) .* ( xOld - xOld1 );
    gm = ones( length( xT ), 1 );
    [ gm( tmp > 0 ), gm( tmp < 0 ) ] = deal( ocPar( 3 ), ocPar( 2 ) );
    asyLU = xT + gm .* [ -( xOld - asyLU( :, 1 ) ), ( asyLU( :, 2 ) - xOld ) ];
end
xL = max( 0.9 * asyLU( :, 1 ) + 0.1 * xT, xL );         % adaptive lower bound
xU = min( 0.9 * asyLU( :, 2 ) + 0.1 * xT, xU );         % adaptive upper bound
% Split (+) and (-) parts of the objective and constraint derivatives -----
[ p0_0, q0_0, p1_0, q1_0 ] = deal( (dg0>0).*dg0, (dg0<0).*dg0, (dg1>0).*dg1, (dg1<0).*dg1 );
[ p0, q0 ] = deal( p0_0 .* ( as( :, 2 ) - xT ).^2, -q0_0 .* ( xT - as( :, 1 ) ).^2 );
[ p1, q1 ] = deal( p1_0 .* ( as( :, 2 ) - xT ).^2, -q1_0 .* ( xT - as( :, 1 ) ).^2 );
% Define the primal and dual projection maps ------------------------------
primalProj = @( lm ) min( xU, max( xL, ( sqrt( p0 + lm * p1 ) .* as( :, 1 ) + ...
sqrt( q0 + lm * q1 ) .* as( :, 2 ) ) ./ ( sqrt( p0 + lm * p1 ) + sqrt( q0 + lm * q1 ) ) ) );
psiDual = @( lm ) g1 - ( ( as( :, 2 ) - xT )' * p1_0 - ( xT - as( :, 1 ) )' * q1_0 ) + ...
sum(p1./(max( as(:,2) - primalProj(lm), 1e-14 ) ) + q1./(max( primalProj(lm) - as(:,1), 1e-14 )));
% Compute the Lagrange multiplier and update design variables -------------
[ lmUp, x, lmid ] = deal( 1e6, xT, -1 );
if psiDual( 0 ) * psiDual( lmUp ) < 0  % check if LM is within the interval
    lmid = fzero( psiDual, [ 0, lmUp ] );
    x = primalProj( lmid );
elseif psiDual( 0 ) < 0                       % constraint cannot be active
   lmid = 0;
   x = primalProj( lmid );
elseif psiDual( lmUp ) > 0                 % constraint cannot be fulfilled
   lmid = lmUp;
   x = primalProj( lmid );
end
end
\end{lstlisting}

\subsection{Main scripts}
 \label{App:CodeListingsMainScripts}

The main scripts \texttt{cshapeTMC.m}, and \texttt{topTMC.m}, which can be used to replicate the results in \autoref{sSec:analysisCshape} and \autoref{sSec:topologyOptimizationExample} are provided in \autoref{lst:cshapeTMC} and \autoref{lst:topTMC}, respectively. The implementation builds upon the framework introduced in \citet{andreassen-etal_11a_88LinesTopOpt} and \citet{ferrari-sigmund_20b_99LinesNewGeneration}. The whole domain $\Omega$ is discretized with a grid of \texttt{nEl=nelx*nely} elements, each of which is assigned a relative density collected in the array \texttt{xPhys}. The element and nodes sorting (\texttt{elNrs}, \texttt{ndNrs}) and DOFs connectivity (\texttt{cDofMat}) follow the convention given in Figure 2 of \cite{andreassen-etal_11a_88LinesTopOpt}. We remark that the shortcuts devised in \cite{ferrari-sigmund_20b_99LinesNewGeneration}, for taking advantage of the matrix symmetry and fast sparse assembly are here ignored, to keep the implementation general.

In the analysis code \texttt{cshapeTMC.m}, the user can specify the domain dimensions (\texttt{Lx}, \texttt{Ly}) and mesh resolution (\texttt{nelx}, \texttt{nely}). Even if the geometry in \autoref{sSec:analysisCshape} is defined by just one length, we give the user the freedom to specify the domain's size and discretization in both directions. For the sake of uniformity with previous codes, the material properties are specified in terms of engineering constants: Young's modulus of solid and voids (\texttt{E0}, \texttt{Emin}), and the Poisson's ratio (\texttt{nu}). These are then converted to the Lam\'{e} parameters ($\lambda,\mu$) within the function \texttt{initializeFEA.m}. Other user-defined parameters, stored in the data structure \texttt{nlP}, define the behaviour of the nonlinear solver: the maximum load multiplier (\texttt{lambdaMax}), number of load increments (\texttt{nIncr}), maximum number iterations (\texttt{maxIter}) and convergence tolerance (\texttt{tolRelRes}) for the Newton solver.

The geometry and mechanical boundary conditions of the C-shape are defined between Lines 17 and 27. On Line 15 the script calls the function initializing and storing all discretization and FE operators (\texttt{initializeFEA.m}), and at the end the nonlinear analysis is performed by the function \texttt{solveIncrIter.m} (Line 29). Here, this function is called with only three input arguments. Therefore, the zero displacement will be used as initial guess for the nonlinear analysis (see also \autoref{lst:solver}).

The script performing topology optimization (\texttt{topTMC.m}, see \autoref{lst:topTMC}) follows the same main assumptions. On each element, we now have also a design variable and a value for the intermediate field, collected in the arrays \texttt{x} and \texttt{xTilde}, respectively. Between Lines 14-22 we specify the parameters used in the density-based TO and optimization formulation, such as the allowed volume fraction (\texttt{volfrac}), RAMP penalization factor (\texttt{qRAMP}), minimum filter radius (\texttt{rmin}), initial values of the Heaviside projection parameters (\texttt{etaD}, \texttt{etaB}, \texttt{beta}), and the $\beta$-continuation rule (\texttt{cntBeta}). The setup of the design domain, together with the loads, boundary conditions and of the passive solid and void regions (if any) is between Lines 26-37. In particular, the definition and treatment of passive solid and void regions, by the indicator arrays \texttt{pasS} and \texttt{pasV}, follows the approach of \cite{ferrari-sigmund_20b_99LinesNewGeneration}. The setup in \autoref{lst:topTMC} can be readily used to replicate the example sketched in \autoref{fig:exampleTopOpt}, giving the outcome of \autoref{fig:TopOptResults1}(iii).

The density filter is implemented following the PDE-based approach (Lines 39-52), as we found it to be more robust than the convolution based operator. Filtering, Heaviside projection, RAMP interpolation, and continuation of penalization/projection parameters are carried out by using anonymous functions, which are defined between Lines 54-61.

The design variables field is initialized to $\rho = 0.5$ on the whole active design domain $\Omega_{D}$, and on Line 70 the nonlinear analysis is carried out for this configuration, up to the end load magnitude $\|\Lambda_{\rm end}\mathbf{f}_{0}\|$. With this setup, we could solve the whole load incremental process with 200 load increments, achieving stable convergence and a highly precise value of the end displacement $\boldsymbol{u}_{\rm end}$.

The iterative re-design loop is performed between Lines 72-99, and involves the usual main steps. The forward analysis is carried out on Line 78, by solving the nonlinear equilibrium equation for the last load step only, using the last computed equilibrium displacement as initial guess. The adjoint analysis is performed on Lines 81-83, the end compliance sensitivity is computed on Line 84, and the chain rule is applied on Lines 83-84. The design update is computed on Line 93-94 by using the simplified MMA-like scheme detailed in \cite{ferrari-etal_21a_250LinesTopOptBuckling} and reported in \autoref{lst:ocUpdate}. In the few last lines, the history variables are updated, some optimization information are printed, and the $\beta$-continuation is applied.

We caution that the present code is not meant for general robustness. In particular, changes to the initial density, or to the end load magnitude will likely require adjustments in the number of load steps to ensure convergence. Also, density filtering affects the stability of the forward analysis in the optimization proceedings, with larger $r_{\rm min}$ values promoting convergence; however, at the price of stiffening the regions intended to be void, thus reducing the accuracy of the contact modelling. Users seeking relevant changes to this basic code, while keeping robustness, may refer to the load-stepping and line-search procedures suggested by \citet{frederiksen-etal_25a_inprovedContact3DTO}, or switch to general-purpose nonlinear solvers.

 \begin{lstlisting}[basicstyle=\footnotesize\ttfamily,breaklines=true,numbers=left,frame=single,
 caption={Load-controlled incremental process, with Newton equilibrium iterations}
 \label{lst:solver}]
function U = solveIncrIter( xPhys, xRAMP, nlP, feP, varargin )
% Initialize arrays and constants used in the iteration -------------------
if nargin > 4, U = varargin{ 1 }; else, U = 0 * feP.F0; end
Lambda = linspace( nlP.lambdaMax / nlP.nIncr, nlP.lambdaMax, nlP.nIncr );
for lStep = 1 : nlP.nIncr % LOOP ON THE LOAD INCREMENT >>>>>>>>>>>>>>>>>>>>
    fprintf( 'Load Step: %1i Load Multiplier: %3.3f \n', lStep, Lambda( lStep ) );
    for iter = 1 : nlP.maxIter % INNER ITERATION ENFORCING EQUILIBRIUM >>>>
        % Assemble tangent stiffness and internal force vector ------------
        [ Kt, Fint ] = assembleKtFi( feP, U( feP.cDofMat )', xRAMP, 'fw', 0 );
        % Compute residual (out-of-balance forces) ------------------------
        rVec = Lambda( lStep ) * feP.F0 - Fint;
        % Check convergence at current Newton step & print information ----
        rrNorm = norm( rVec( feP.free ) ) / norm( Lambda( lStep ) * feP.F0 );
        fprintf( 'Newton It.: %1i, ||RelRes||_2: %0.3e \n', iter, rrNorm );
        if rrNorm <= nlP.tolRelRes, break; end
        % Update displacement vector by Newton correction -----------------
        U( feP.free ) = U( feP.free ) + Kt( feP.free, feP.free ) \ rVec( feP.free );
    end
    % Check convergence failure -------------------------------------------
    if iter == nlP.maxIter && rrNorm > nlP.tolRelRes
        error( 'Solver failed to converge // RelResNorm : %0.3e \n', rrNorm );
    end
    % Plot current deformed configuration ---------------------------------
    plotDeformed( feP.X, feP.IX, xPhys, U );
end
\end{lstlisting}

\subsection{Nonlinear Equilibrium Solver}
 \label{sApp:incrementalIterativeSolver}

The incremental-iterative analysis is implemented in the function in \autoref{lst:solver}. At each load step, with external loading $\Lambda_{i}\mathbf{f}_{0}$, the equilibrium is restored by solving \autoref{eq:variationTPE} by the Newton-Raphson method.

In the $k$-th Newton step, the displacement correction $\Delta\mathbf{u}^{(k)}$ is computed by solving the FE-discretized algebraic equations
\begin{equation}
 \label{eqdiscretizedNewtonStep}
  \mathbf{K}_{T}(\mathbf{u}^{(k)})
  \Delta\mathbf{u}^{(k)} = \Lambda_{i}\mathbf{f}_{0} - \mathbf{f}_{\rm int}(\mathbf{u}^{(k)})
\end{equation}
until $\|\Lambda_{i}\mathbf{f}_{0} - \mathbf{f}_{\rm int}(\mathbf{u}^{(k)})\|_{2}/\|\mathbf{f}_{0}\|_{2} \leq \tau$. The tangent stiffness matrix $\mathbf{K}_{T}$ and the internal forces $\mathbf{f}_{\rm int}$ are computed by the routine described in the following subsection. The function requires a minimum of three input arguments; a fourth one is used to set the initial guess for the equilibrium displacement, used by the Newton process at the first load step.

Here we consider a simple load controlled procedure, and a Newton scheme with no damping or line search. Thus, arbitrary changes in the parameters or in the test example may hamper the robustness of the solver, in which case the user may extend the current code with more robust path-following procedures and/or nonlinear solvers.

 \begin{lstlisting}[basicstyle=\footnotesize\ttfamily,breaklines=true,numbers=left,frame=single,
caption={Assembly of the tangent stiffness matrix and internal force vector}
 \label{lst:assembleKtFi}]
function [ Kt, Fint, Psi ] = assembleKtFi( feP, ue, xPhys, aType, dxPhys )
% Initialize relevant dimensions and corresponding arrays -----------------
nEl = size( feP.cDofMat, 1 );                          % number of elements
nQp = length( feP.wG( : ) );      % number of quadrature points per element
iF = zeros( 4, nQp, nEl ); Aq = zeros( 3, 4, nQp, nEl );
Cmat = zeros( 3, 3, nQp, nEl ); Ts = zeros( 4, 4, nQp, nEl );
% Scaling of the material constants by the relative density ---------------
xPhys  = reshape( xPhys( : ), 1, 1, 1, [] );
lam    = feP.lam .* xPhys;                          % First Lame' parameter
mu     = feP.mu .* xPhys;          % Second Lame' parameter (shear modulus)
% Compute displacement gradient on each quadrature point ------------------
dudx = squeeze( feP.gradN( 1, :, : ) )' * ue( 1 : 2 : end, : );     % du/dx
dvdx = squeeze( feP.gradN( 1, :, : ) )' * ue( 2 : 2 : end, : );     % dv/dx  
dudy = squeeze( feP.gradN( 2, :, : ) )' * ue( 1 : 2 : end, : );     % du/dy
dvdy = squeeze( feP.gradN( 2, :, : ) )' * ue( 2 : 2 : end, : );     % dv/dy
tmp0 = [ nQp, 1, nEl ];
gradU = cat( 2, reshape(dudx,tmp0), reshape(dvdx,tmp0), reshape(dudy,tmp0), reshape(dvdy,tmp0) );
% Deformation gradient F = I + grad( u ), its determinant and inverse -----
F  = cat( 2, reshape(1+dudx,tmp0), reshape(dvdx,tmp0), reshape(dudy,tmp0), reshape(1+dvdy,tmp0) );
dF = reshape( F(:,1,:) .* F(:,4,:) - F(:,2,:) .* F(:,3,:), 1, 1, nQp, nEl );
F  = reshape( permute( F, [ 2, 1, 3, 4 ] ), [ 2, 2, nQp, nEl ] );
iF( 1, :, : ) =  F( 2, 2, :, : ) ./ dF;
iF( 2, :, : ) = -F( 1, 2, :, : ) ./ dF;
iF( 3, :, : ) = -F( 2, 1, :, : ) ./ dF;
iF( 4, :, : ) =  F( 1, 1, :, : ) ./ dF;
% Build nonlinear strain-displacement matrix B_1 = B_0 + A_q * G ----------
Aq( 1, 1 : 2, :, : ) = permute( gradU( :, 1 : 2, : ), [ 4, 2, 1, 3 ] );
Aq( 2, 3 : 4, :, : ) = permute( gradU( :, 3 : 4, : ), [ 4, 2, 1, 3 ] );
Aq( 3, 1 : 2, :, : ) = permute( gradU( :, 3 : 4, : ), [ 4, 2, 1, 3 ] );
Aq( 3, 3 : 4, :, : ) = permute( gradU( :, 1 : 2, : ), [ 4, 2, 1, 3 ] );
B1 = feP.B0( :, :, : ) + pagemtimes( Aq, feP.G( :, :, : ) );
% Compute right Cauchy-Green tensor rCG = F' * F and its inverse ----------
rCG = pagemtimes( pagetranspose( F ), F );
det_rCG = rCG( 1, 1, :, : ) .* rCG( 2, 2, :, : ) - rCG( 1, 2, :, : ).^2;
irCG = rCG .* 0;
irCG( 1, 1, :, : ) =  rCG( 2, 2, :, : ) ./ det_rCG;
irCG( 1, 2, :, : ) = -rCG( 1, 2, :, : ) ./ det_rCG;
irCG( 2, 1, :, : ) = -rCG( 2, 1, :, : ) ./ det_rCG;
irCG( 2, 2, :, : ) =  rCG( 1, 1, :, : ) ./ det_rCG;
% II Piola-Kirchhoff stress (S = lam*ln(J)*C^-1 + mu*(I-C^-1)) ------------
sPK2 = lam .* log( dF ) .* [ irCG( 1, 1, :, : ); irCG( 2, 2, :, : ); irCG( 1, 2, :, : ) ] + ...
    mu .* [ 1 - irCG( 1, 1, :, : ); 1 - irCG( 2, 2, :, : ); -irCG( 1, 2, :, : ) ];
% Material tangent tensor (4th order elasticity tensor, Voigt notation) ---
t0   = mu - lam .* log( dF );
Cmat(1,1,:,:) = ( lam + 2 * t0 ) .* irCG( 1, 1, :, : ).^2;
Cmat(2,1,:,:) = lam .* irCG(1,1,:,:) .* irCG( 2, 2, :, : ) + 2 * t0 .* irCG( 1, 2, :, : ).^2;
Cmat(3,1,:,:) = lam .* irCG(1,1,:,:) .* irCG(1,2,:,:) + 2 * t0 .* irCG(1,1,:,:) .* irCG(1,2,:,:);
Cmat(2,2,:,:) = ( lam + 2 * t0 ) .* irCG( 2, 2, :, : ).^2;
Cmat(3,2,:,:) = lam .* irCG(2,2,:,:) .* irCG(1,2,:,:) + 2 * t0 .* irCG(1,2,:,:) .* irCG(2,2,:,:);
Cmat(3,3,:,:) = lam .* irCG(1,2,:,:).^2 + t0 .* ( irCG(1,1,:,:) .* irCG(2,2,:,:) + irCG(1,2,:,:).^2 );
Cmat = Cmat + pagetranspose( Cmat ) - Cmat .* eye( 3 );
% Geometric stiffness contribution (stress tensor in matrix form) ---------
Ts( 1, 1, :, : ) = sPK2( 1, :, :, : );
Ts( 2, 2, :, : ) = sPK2( 1, :, :, : );
Ts( 1, 3, :, : ) = sPK2( 3, :, :, : );
Ts( 2, 4, :, : ) = sPK2( 3, :, :, : );
Ts( 3, 1, :, : ) = sPK2( 3, :, :, : );
Ts( 4, 2, :, : ) = sPK2( 3, :, :, : );
Ts( 3, 3, :, : ) = sPK2( 2, :, :, : );
Ts( 4, 4, :, : ) = sPK2( 2, :, :, : );
% Ccurrent value of the strain energy density -----------------------------
Psi = 0.5 * ( lam .* log( dF ).^2 + mu .* ( rCG(1,1,:,: ) + rCG(2,2,:,:) - 2 ) ) - mu .* log( dF );
Psi = sum( squeeze( Psi ), 1 )';
% Build the HuHu regularization contribution to the tangent stiffness -----
HxH  = pagemtimes( pagetranspose( feP.Hkron ), feP.Hkron );
tmp0 = pagemtimes( HxH, ue );
tmp1 = pagemtimes( permute( permute( iF, [ 1 3 2 ] ), [ 2 1 3 ] ), feP.G );
tmp2 = permute( reshape( tmp0, size( tmp0, 1 ), size( tmp0, 2 ), 1, size(tmp0, 3 ) ) .* ...
    reshape( tmp1, 1, size( tmp1, 1 ), size( tmp1, 2 ), size( tmp1, 3 ) ), [ 1 3 4 2 ] );             
% Element tangent stiffness matrix keT = int(B1'*C*B1 + G'*Ts*G + HuHu) ---
wGr  = reshape( feP.wG( : ), 1, 1, nQp, 1 );
HuHu = feP.kr * exp( -5 * dF ) .* ( HxH - 5 .* dF .* tmp2 );
keT  = sum( wGr .* ( pagemtimes( pagemtimes( pagetranspose( B1 ), Cmat ), B1 ) + ...
    pagemtimes( pagemtimes( pagetranspose( feP.G ), Ts ), feP.G ) + HuHu ), 3 );
% Assemble global tangent stiffness matrix --------------------------------
keT = reshape( keT, feP.nElDof * feP.nElDof, nEl )';
Kt = sparse( feP.iK, feP.jK, reshape( keT', [], 1 ), feP.nDof, feP.nDof );
% Compute internal force vector based on the analysis type ----------------
if strcmp( aType, 'fw' )                     % forward equilibrium analysis
    feI = sum( wGr .* ( pagemtimes( pagetranspose( B1 ), sPK2 ) + ...
        feP.kr * exp( -5 * dF ) .* pagemtimes( HxH, reshape( ue, 8, 1, 1, nEl ) ) ), 3 );
    [ ind, sz ] = deal( feP.cDofMat( : ), 1 );
    feI = reshape( feI, feP.nElDof, nEl )';
elseif strcmp( aType, 'ad' )                 % adjoint sensitivity analysis
    dxPhys = reshape( dxPhys( : ), 1, 1, 1, [] );
    [ dlam, dmu ] = deal( feP.lam .* dxPhys, feP.mu .* dxPhys );
    % Sensitivity of stress with respect to design field
    dsPK2 = dlam .* log( dF ) .* [ irCG( 1, 1, :, : ); irCG( 2, 2, :, : ); irCG( 1, 2, :, : ) ] + ...
        dmu .* [ 1 - irCG( 1, 1, :, : ); 1 - irCG( 2, 2, :, : ); -irCG( 1, 2,:, : ) ];
    feI = sum( wGr .* pagemtimes( pagetranspose( B1 ), dsPK2 ), 3 );
    feI = reshape( feI, feP.nElDof, nEl );
    [ ind, sz ] = deal([reshape(feP.cDofMat',[],1), reshape(repelem((1:nEl)',feP.nElDof),[],1)],nEl);
end
Fint = accumarray( ind, feI( : ), [ feP.nDof, sz ] );
end
\end{lstlisting}

\subsection{Assembly of the tangent stiffness matrix and internal force vector}
 \label{sApp:assemblyKtFi}

At each step of the incremental-iterative process, the routine in \autoref{lst:assembleKtFi} computes the elemental tangent matrices $\mathbf{k}_{T(e)}$ and internal force vector $\mathbf{f}_{{\rm int}(e)}$, and assembles these in the global arrays.

The internal forces and tangent stiffness are given by the first and second variations of the SED, respectively, which considering the \textsf{HuHu} regularization read
\begin{align}
 \label{eq:firstVariationSED_HuHu_App}
 \delta\tilde{W}(\boldsymbol{u};\delta\boldsymbol{u}) & =
   \boldsymbol{P}\cdot\delta\boldsymbol{F} 
   + k_{r} e^{-5|\boldsymbol{F}|}
  \mathbb{H}\boldsymbol{u}
 \cdot \mathbb{H}\delta\boldsymbol{u} \\
 \label{eq:secondVariationSED_HuHu}
 \delta^{2}\tilde{W}(\boldsymbol{u} ; \delta\boldsymbol{u}) & = 
 \delta\boldsymbol{P}\cdot\delta\boldsymbol{F} +
 \boldsymbol{P}\cdot\delta^{2}\boldsymbol{F} +
 k_{r}e^{-5|\boldsymbol{F}|}
 \left(
 \mathbb{H}\delta\boldsymbol{u}
 \cdot\:\mathbb{H}\delta\boldsymbol{u}
 - 5|\boldsymbol{F}|^{2}
 {\rm tr}(\boldsymbol{F}^{-1}\delta\boldsymbol{F})\mathbb{H}\delta\boldsymbol{u}
 \cdot\:\mathbb{H}\boldsymbol{u}
 \right)
\end{align}

Since we adopt a Total Lagrangian approach based on the Green-Lagrange strain $2\boldsymbol{\gamma} = \boldsymbol{F}^{T}\boldsymbol{F} - \boldsymbol{I}$, the stress tensor $\boldsymbol{P}$ is transformed to the second Piola-Kirchhoff stress tensor $\boldsymbol{S} = \lambda \ln|\boldsymbol{F}|\boldsymbol{F}^{-1}\boldsymbol{F}^{-T} + \mu (\boldsymbol{I}-\boldsymbol{F}^{-1}\boldsymbol{F}^{-T})$.

Upon FE discretization,\eqref{eq:firstVariationSED_HuHu_App}-\eqref{eq:secondVariationSED_HuHu} give the vector of internal forces and the tangent stiffness matrix, consisting of material, geometric, and stabilization contributions
\begin{align}
 \label{eq:discretizedInternalForceVector}
 \mathbf{f}_{\rm int(e)} & = \int_{\Omega_e} \mathbf{B}^T_{\gamma} \mathbf{T}(\boldsymbol{S}) \: {\rm d}\Omega +
 k_{r}\int_{\Omega_e} e^{-5|\mathbf{F}|} \mathbf{H}^T\mathbf{H}\mathbf{u}
 \: {\rm d}\Omega \\
 \label{eq:discretizedTangentStiffness}
 \mathbf{k}_{T(e)} & = \int_{\Omega_e} \mathbf{B}^T_{\gamma} \mathbf{C} \mathbf{B}_{\gamma} + \mathbf{G}^T \mathbf{T}(\boldsymbol{S}) \mathbf{G} \: {\rm d}\Omega +
 k_{r}\int_{\Omega_e} e^{-5|\mathbf{F}|} \left(\mathbf{H}^T\mathbf{H} - 5|\mathbf{F}|(\mathbf{F}^{-T}\mathbf{G})\otimes(\mathbf{H}
 \mathbf{u})\right) \: {\rm d}\Omega
\end{align}
where $\mathbf{B}_{\gamma} = \mathbf{B}_{0} + \mathbf{A}(\mathbf{u})\mathbf{G}$ is the nonlinear strain-displacement matrix, $\mathbf{C}$ is the matrix collecting the material tangent moduli, and $\mathbf{T}(\boldsymbol{S})$ represents the stress stiffness contribution \citep{book:crisfield91}.

The last contribution in \eqref{eq:discretizedTangentStiffness}, depending on the inverse of the deformation gradient $\mathbf{F}^{-1}$, generally makes the tangent matrix nonsymmetric. Even if this contribution appears to be relatively small, in the whole \textsf{HuHu}-regularization term, we observed that it plays a crucial role in speeding up convergence of the Newton iterations, and in making the whole nonlinear solve more robust (see \autoref{sSec:analysisCshape}).

Finally, we highlight that the same routine is also used to compute the right hand side for the adjoint problem \eqref{eq:endComplianceSensitivityContinuous}, which is needed for performing the sensitivity analysis.

 \begin{lstlisting}[basicstyle=\footnotesize\ttfamily,breaklines=true,numbers=left,
frame=single,caption={Computation of discretization arrays and FE operators}
 \label{lst:initializeFEA}]
function feP = initializeFEA( Lx, Ly, nelx, nely, E0, nu, alpha )
%% Compute quantities which are constant in the Finite Element Analysis >>>
nNdEl  = 4;            % number of nodes per element (4-node quadrilateral)
nElDof = 2 * nNdEl;                                % total DOFs per element
nEl    = nelx * nely;                % total number of elements in the mesh
% Create element and node numbering ---------------------------------------
elNrs = reshape( 1 : nEl, nely, nelx );
ndNrs = reshape( 1 : ( 1 + nelx ) * ( 1 + nely ), 1 + nely, 1 + nelx );
% Create DOF connectivity matrix for each element -------------------------
cDofMat = reshape(2*ndNrs(1:end-1,1:end-1)+1,nEl,1)+[0,1,2*nely+[2,3,0,1],-2,-1];
nDof = max( cDofMat( : ) );
% Generate node coordinates in physical space (useful for plotting) -------
[ xx, yy ] = meshgrid( linspace( 0, Lx, nelx+1 ), linspace( 0, -Ly, nely+1 ) );
X  = [ xx( : ), -flipud( yy( : ) ) ];                   % coordinate matrix
IX = reshape(ndNrs(1:end-1,1:end-1)+1,nEl,1)+[0,nely+[1,0],-1]; % nodes connectivity
% Precompute index pairs for global stiffness matrix assembly -------------
iK = reshape( kron( cDofMat, ones( nElDof, 1 ) )', nElDof^2 * nEl, 1 );
jK = reshape( kron( cDofMat, ones( 1, nElDof ) )', nElDof^2 * nEl, 1 );
% Quadrature points and weights (Gauss-Lobatto 3x3 quadrature rule) -------
[ xiG, etaG ] = meshgrid( [ -1, 0, 1 ], [ -1, 0, 1 ] );
wG = kron( [ 1, 4, 1 ], [ 1, 4, 1 ] ) ./ 9;
% Shape function derivatives with respect to reference coordinates (xi,eta)
dN = @( xi, eta ) [ eta-1, 1-eta, 1+eta, -1-eta; xi-1,-1-xi, 1+xi, 1-xi ] / 4;
% Second derivatives used in the HuHu regularization (3rd index is node) --
d2N( :, :,   1 ) = [ 0, 1 ; 1, 0 ] / 4;
d2N( :, :,   2 ) = - d2N( :, :,   1 );
d2N( :, :, 3:4 ) =   d2N( :, :, 1:2 );
% Define physical coordinates of the reference element corners ------------
xp = [ 0, 0; Lx/nelx, 0; Lx/nelx, Ly/nely; 0, Ly/nely ];
% Precompute element-wise matrices for each quadrature point --------------
for j = 1 : length( wG( : ) )
    % ------------------ element Jacobian, Jacobian determinant and inverse
    iJ = ( dN( xiG( j ), etaG( j ) ) * xp ) \ eye( 2 );
    % -------------------- physical gradient: d_(x,y)N = J^-1 * d_(xi,eta)N
    gradN( :, :, j ) = iJ * dN( xiG( j ), etaG( j ) );
    % ------------ physical Hessian with H_(x,y) = J^-1 * H_(xi,eta) * J^-T
    hessN( :, :, j ) = reshape( pagemtimes( iJ, pagemtimes( d2N, iJ' ) ), [], 4 );
    % ------------- Kronecker product for higher-order regularization terms
    Hkron( :, :, j ) = kron( hessN( :, :, j ), eye( 2 ) );
    % ---------------- scale integration weight by the Jacobian determinant
    wG( j ) = wG( j ) / det( iJ );
    % ----------------- strain-displacement matrix at this quadrature point
    G( :, :, j )  = kron( gradN( :, :, j ), eye( 2 ) );
    B0( :, :, j ) = [ 1, 0, 0, 0; 0, 0, 0, 1; 0, 1, 1, 0 ] * G( :, :, j );    
end
% Transform the material and regularization parameters --------------------
lam    = E0 * nu / ( ( 1 + nu ) * ( 1 - 2 * nu ) );   % Lame' 1st parameter
mu     = E0 / ( 2 * ( 1 + nu ) );                     % Lame' 2nd parameter
kappas = E0 / ( 3 * ( 1 - 2 * nu ) );           % bulk modulus on the solid
kr     = alpha * Lx^2 * ( kappas + 4/3*mu );   % HuHu regularization factor
% Store useful information in the data structure "feP" --------------------
feP = struct('nEl', nEl, 'nElDof', nElDof, 'cDofMat', cDofMat, 'nDof', nDof, 'iK', iK, 'jK', jK, ...
    'wG', wG, 'gradN', gradN, 'hessN', hessN, 'Hkron', Hkron, 'B0', B0, 'G', G, 'lam', lam, ...
    'mu', mu, 'kappas', kappas, 'kr', kr, 'elNrs', elNrs, 'ndNrs', ndNrs, 'X', X, 'IX', IX);
end
\end{lstlisting}

\subsection{Discretization and FE operators initialization}

All finite element operators which are design- and displacement-independent are set up in the routine \texttt{initializeFEA.m}, and stored in the data structure \texttt{feP}. 

First, the parameters and arrays representing the discretization (\texttt{nNdEl}, \texttt{nElDof}, \texttt{nEl}, \texttt{elNrs}, \texttt{ndNrs}, \texttt{nDof}, \texttt{X},\texttt{IX}), and DOFs connectivity (\texttt{cDofMat}, \texttt{iK}, \texttt{jK}) are defined between Lines 3-18, following the mesh sorting and connectivity conventions adopted by \cite{andreassen-etal_11a_88LinesTopOpt}.

Then, the operators discretizing the deformation gradient ($\mathbf{G}$), the linear strain-displacement matrix ($\mathbf{B}_{0}$), and the displacement Hessian ($\mathbf{H}$), which depend neither on the design ($\rho$), nor on the displacement variables ($\mathbf{u}$), are computed only once.

To this end we first define the arrays collecting the shape functions' first and second derivatives w.r.t. the logical coordinates $(\xi,\zeta)$: $\nabla_{(\xi, \zeta)}\mathbf{N}$ and $\nabla^{2}_{(\xi, \zeta)}\mathbf{N}$, where $\mathbf{N}(\xi,\zeta) = [ N_{1}(\xi,\zeta), N_{2}(\xi,\zeta), \dots, N_{k}(\xi,\zeta)]$ (Lines 23-27). Then, between Lines 31-45 the following operators are computed
\begin{equation}
 \begin{aligned}
 \label{eq:GB1definition}
  \mathbf{G} & = \left[
  \nabla_{(x, y)}N_{i} \otimes I_{2}
  \right]_{i=1, \ldots, k} \\
  \mathbf{B}_{0} & = \mathbf{L}\mathbf{N} \\
  \mathbf{H} & = \left[
  \nabla^{2}_{(x, y)}H_{i} \otimes 
  I_{2} \right]_{i=1,\ldots, k}
 \end{aligned}
\end{equation}
where $\nabla_{(x, y)}N_{i} = \mathbf{J}^{-1} \nabla_{(\xi, \zeta)}N_{i}$, $\nabla^{2}_{(x,y)}N_{i} = \mathbf{J}^{-1}\nabla^{2}_{(\epsilon,\eta)}N_{i} \mathbf{J}^{-T}$, $\mathbf{L}$ is the strain-displacement combination matrix, and ``$\otimes$'' denotes the dyadic (Kronecker) product of two arrays. 

The operators in \eqref{eq:GB1definition} are computed at each of the quadrature point introduced in Line 20, and therefore are stored in three-dimensional arrays (i.e., $G(i,j,k) = \left. G_{ij} \right|_{(\xi_{k},\eta_{k})}$). Thus, the code can be extended to higher order elements, upon changing to the corresponding shape function derivatives.  However, we caution that for higher order elements, and for unstructured $\mathcal{Q}_{1}$ discretizations, the higher-order derivatives of the isoparametric mapping ($\xi_{xx}$, $\xi_{xy}$, $\xi_{yy}$, $\eta_{xx}$, $\eta_{xy}$, $\eta_{yy}$) should be taken into account. In our implementation, these are neglected as we refer to a structured discretization of bilinear $\mathcal{Q}_{1}$ elements.

Finally, the routine computes the material parameters ($\lambda, \mu, k_{s}, k_{r}$) starting from the input data ($E_{0}, \nu, \alpha$) following well-known elasticity relationships (Lines 47-50)
\begin{equation}
 \lambda = \frac{\nu E}
  {(1+\nu)(1-2\nu)} \: , \qquad
  \mu = \frac{E}{2(1-\nu)}
\end{equation}
which are valid for a compressible solid in state of plane strain.

 \begin{lstlisting}[basicstyle=\footnotesize\ttfamily,breaklines=true,numbers=left,frame=single, 
caption={Routine plotting a field (\texttt{fld}) over the deformed mesh}
\label{lst:plotDeformed}]
function plotDeformed( X, IX, fld, U )
clf reset; ax = gca;
% Create colored patch plot on the deformed geometry ----------------------
patch( ax, 'Faces', IX, 'Vertices', [ X(:,1)+U(1:2:end), X(:,2)+U(2:2:end) ], ...
'FaceVertexCData', fld, 'FaceColor', 'flat', 'EdgeColor', 'blue', 'LineWidth', .5, 'Marker','none');
% Set color axis limits, colorbar and background appearance ---------------
clim( [ min( fld ), max( fld ) ] ); axis off equal; colormap( flipud( gray ) );
clb = colorbar(); clb.FontSize = 16; clb.Location = 'eastoutside';
fig = gcf(); fig.Color = [ 1, 1, 1 ]; drawnow;
end
\end{lstlisting}

\subsection{Plotting subroutine}
\label{sec:visualization}

The field of relative densities $\hat{\rho}(\boldsymbol{x})$ is visualized on the deformed mesh using the function in \autoref{lst:plotDeformed}. The deformed coordinates are computed as $\mathbf{x} = \mathbf{x}_0 + \mathbf{u}$, where $\mathbf{x}_0$ collects the coordinates of the reference domain and $\mathbf{u}$ is a given displacement field. The routine can be further used, with minor modifications, also to plot any other element-based field.

\end{appendices}
\end{document}